\documentclass[useAMS, usenatbib]{mnras}

\usepackage{graphicx}

\usepackage[usenames]{color}
\usepackage{colortbl}

\title[Testing stellar proper motions of the TGAS using data of the HSOY, UCAC5 and PMA catalogues] {Testing stellar proper motions of the TGAS using data of the HSOY, UCAC5 and PMA catalogues}
\author[P.N. ~Fedorov et al.] {\thanks {E-mail: pnfedorov@gmail.com (PNF); akhmetovvs@gmail.com (AVS); astronomo@mail.ru (VAB)} 
P.N. ~Fedorov$^1$, V.S.~Akhmetov$^1$, A.B.~Velichko$^1$. \\
$^1$Institute of Astronomy V.N. Karazin Kharkiv National University, Sums'ka 35, 61022, Kharkiv, Ukraine. \\}

\begin{document}

\date{Accepted 2017 . Received 2017 ; in original form 2017}

\pagerange{\pageref{firstpage}--\pageref{lastpage}} \pubyear{2017}

\def\LaTeX{L\kern-.36em\raise.3ex\hbox{a}\kern-.15em T\kern-.1667em\lower.7ex\hbox{E}\kern-.125emX}

\newtheorem{theorem}{Theorem}[section]

\maketitle

\label{firstpage}

\begin{abstract}
We present an investigation of stellar proper motions of the TGAS catalogue and ground-based HSOY, UCAC5 and PMA catalogues derived by combining with the Gaia DR1 space data. This investigation concerns with only those stars of ground-based catalogues that are contained in the TGAS.
Supposing that systematic differences of stellar proper motions of two catalogues are caused by mutual solid-body rotation of coordinate systems, which are realized by the catalogues under comparison, we analyze components of the mutual rotation vector between these systems.
It has been found that in all three  cases of comparison of the HSOY, UCAC-5 and PMA catalogues with TGAS, the $\omega_{y}$  component of the mutual rotation vector depends on magnitude nonlinearly within the range $9.5 < m< 11.5$, and has an amplitude reaching 1.5 mas/yr. 
The analysis has shown that the reason causing this effect is presence in proper motions of the Tycho-2-stars containing in the TGAS, of some inexplicable dependency on magnitude. It has been shown that proper motions of the TGAS stars derived using AGIS differ from those derived as a result of application the conventional (classical) method. At the same time, such a difference in proper motions of Hipparcos-stars from TGAS derived by both methods was not found. Investigation of systematic differences between proper motions of the TGAS stars derived by the classical method and proper motions of the HSOY, UCAC5 and PMA stars has shown that the values of the $\omega_{y}$ component in this case do not undergo any jumps and do not depend on magnitude. That fact unambiguously indicates that some magnitude error is contained in proper motions of Tycho-2-stars derived by the use of AGIS. In the framework of the solid-body rotation model the coordinate systems set by the TGAS catalogue with classical proper motions (
on the one hand) and by the HSOY, UCAC5 and PMA catalogues (on the other hand) have mutual rotation with the components along axis less than 0.2-0.3 mas/yr. Average value of modules of angular velocities considered as the measure of residual rotation, do not exceed 0.4 mas/yr. It has been shown that in the range $7.75 < m < 9.75$ proper motions of Hipparcos-stars from the PMA and TGAS catalogues do not differ systematically by more than 0.3 mas/yr and almost free of magnitude equation.

\end{abstract}


\section{Introduction}

The first release of the Gaia mission \citep {b1,f1,l1} marked the beginning of a new era in astrometry. The astronomical community has been looking forward to its first results. Although the first data release \citep {l1} contains preliminary astronomical results based on observations during only the first 14 months of its operational phase, they already now are used  intensively in many astronomical applications. The most often the TGAS catalogue data \citep* {m4} providing positions, proper motions and parallaxes with tipical accuracy of Hipparcos level or better for about 2 million objects up to $\sim$11.5 magnitude  is used.

For instance, during one year three new astrometric catalogues — HSOY \citep {a4}, UCAC5 \citep {z1} and PMA \citep {a5} containing proper motions appeared. All of them used in varying degrees the Gaia DR1 data. When creating the UCAC5 catalogue positions and proper motions of the TGAS stars for reduction the US Naval Observatory CCD Astrograph Catalog (UCAC) observational data to the Gaia DR1 reference frame were used resulting in proper motions for more than 107 million stars with tipical accuracy from 1 to 2 mas/yr in the $11^{m}<R<15^{m}$ magnitude range and near 5 mas/yr at 16-th mag.

The HSOY catalogue were derived by combining positions from the PPMXL \citep {r1} and Gaia DR1 using the weighted least square method which were used in deriving the PPMXL its own. This catalogue contains 583 million stars with positions of Gaia DR1 quality and proper motions accurate to from less than 1 mas/yr to 5 mas/yr depending on magnitude and coordinates of objects in the sky. 

In fact, when creating these two catalogues the classical method of expansion of the reference system from bright part of a magnitude range to the faintest stars was used. The overwhelming majority of catalogues obtained from ground-based observations and containing proper motions were derived with the use of the reference stars of Hipparcos/Tycho-2 \citep{p1,h1}.

Proper motions of the PMA catalogue objects were derived by combining the 2MASS \citep {s2} and Gaia DR1 positions. The catalogue contains about 420 million objects with G-magnitudes from 8 to 21, absolute proper motions and taken from the Gaia DR1 positions. Zero-point of the proper motions was set by positions of about 1.6 extragalactic sources extracted from the Gaia DR1 data by a special procedure. The mean formal error of the absolute calibration is less than 0.35 mas/yr. The RMS error of proper motions depends on stellar magnitude and constitutes from 2-5 mas/yr for stars with $10^{m}<G<17^{m}$ to 5-10 mas/yr for faint ones. The PMA catalogue expands the reference system set by positions of galaxies in faint part of the magnitude range to the bright one unlike the HSOY and UCAC5.

In section 2 we carry out the global (through the whole celestial sphere) comparison of proper motions of sources of the catalogues under consideration. For more detailed investigation we use the model of systematic differences of proper motions beliewing that they are caused by mutual solid-body rotation of the coordinate systems. The model parameters we derive by the least square method and note that proper motions of all catalogues under consideration contain systematic differences from ones of the TGAS catalogue depending on magnitude.

In section 3 we show that the original proper motions (derived in AGIS procedure \citep{l3}) of the TGAS-stars differ from ones derived by the conventional classical method, i.e. calculated as the positional difference between positions of primary sources taken from the Tycho-2 \citep{h1} and Hipparcos \citep{l2} at epoch near J1991.0 and their Gaia DR1 positions at epoch J2015.0, divided by the epoch difference. Also we stress that the differences between original and classical proper motions exist only for Tycho-2 stars but is absent for Hipparcos stars containing in the TGAS.

Then in section 4 using classical proper motions of the TGAS-stars we again analyze differences of proper motions in form of (catalogue-TGAS $_{class}$) and show that the value of the component of mutual rotation vector $\omega_{y}$  practically does not depend on magnitude thus demonstrating existence of systematic error in original proper motions of the TGAS. Here we discuss behaviour of the  $\omega_{z}$  component and assume that its insignificant dependence on madnitude is probably caused by errors in proper motions of the PMA and UCAC5. In discussing the results (section 5) we note that the TGAS proper motions derived using the AGIS procedure have an unexplained inexplicable dependence on magnitude, vanishing when the original proper motions are replaced by the classical ones. Also we show that the systems of proper motions of the HSOY, UCAC5 and PMA do not have systematic differences exceeding 0.5 mas/yr.

\section{Comparison of stellar proper motions}

\subsection{Investigation of proper motions of the PMA-TGAS} 

\begin{figure*}
\vspace*{0pt}
\includegraphics[width = 87mm,]{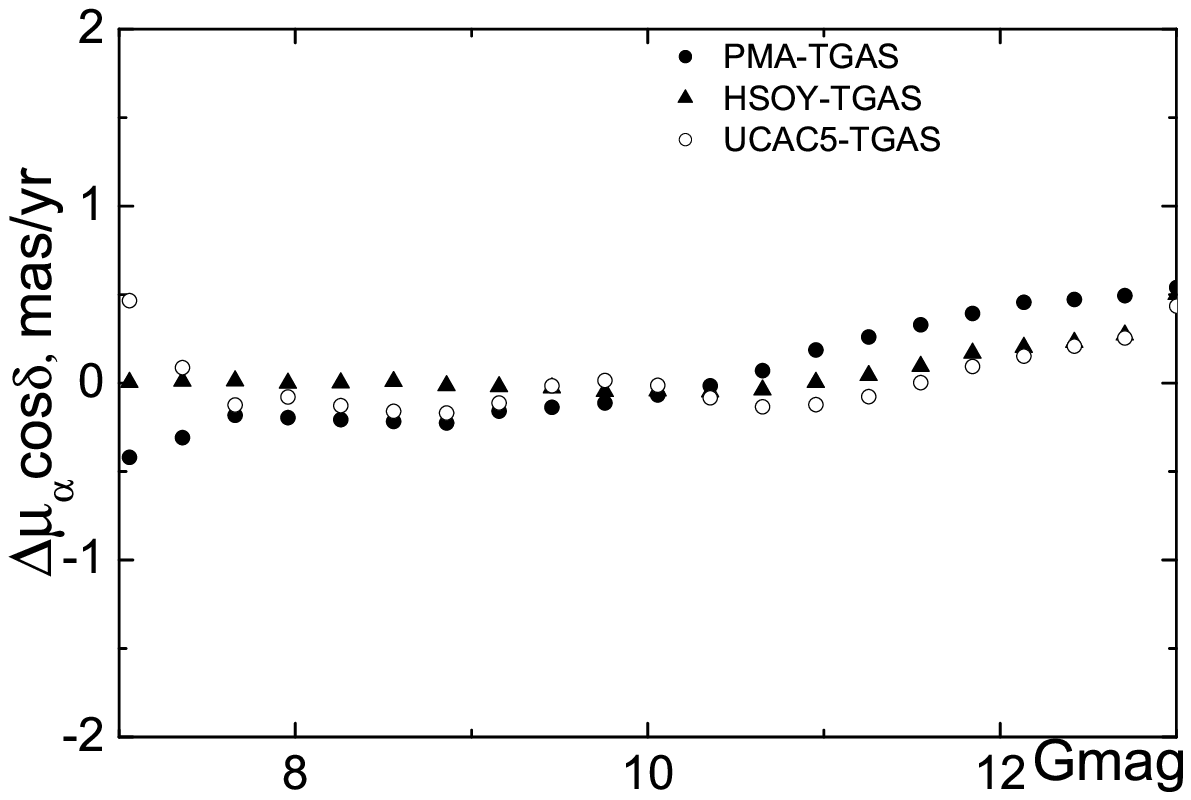}
\includegraphics[width = 87mm,]{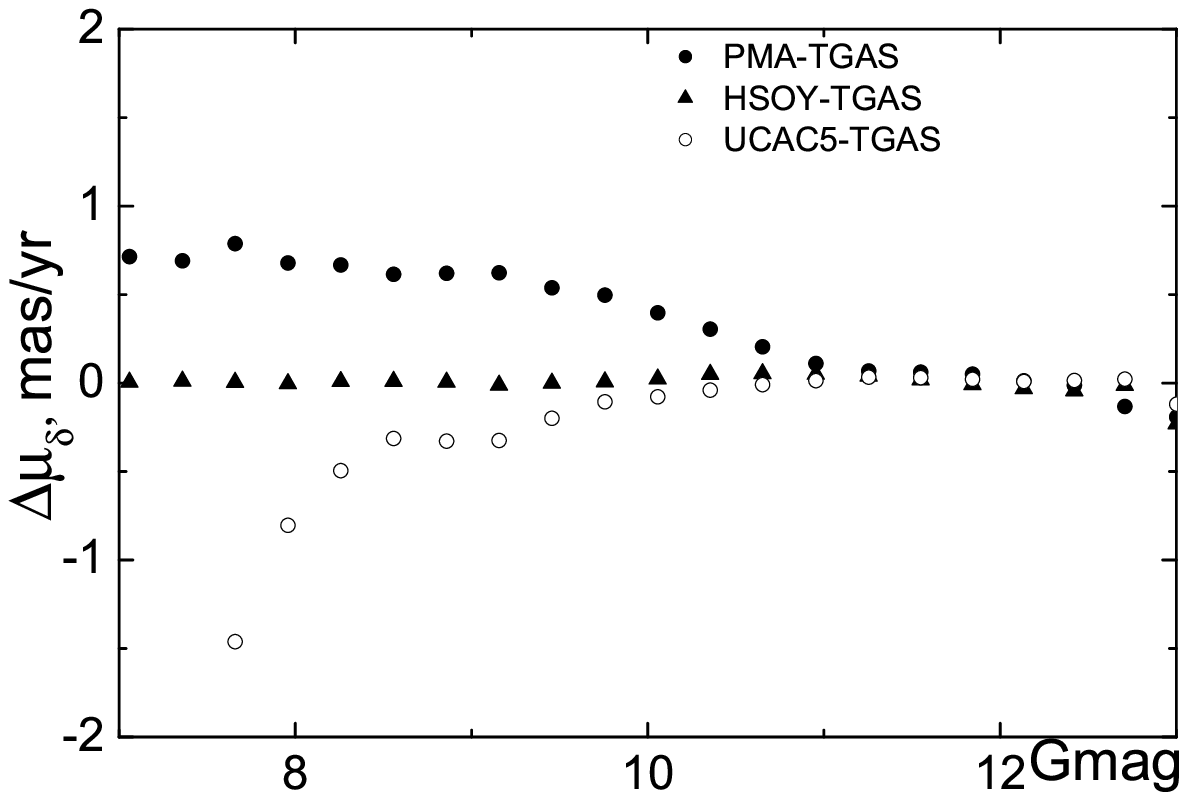}
\caption{Dependencies of stellar proper motion differences  $\bigtriangleup\mu_{\alpha} cos \delta$ (left) and  $\bigtriangleup\mu_{\delta}$ (right) of the PMA-TGAS, HSOY-TGAS and UCAC5-TGAS catalogues as a function of G magnitude.}
\label{m_a_thup}
\end{figure*}

\begin{figure*}
\vspace*{0pt}
\includegraphics[width = 58mm,]{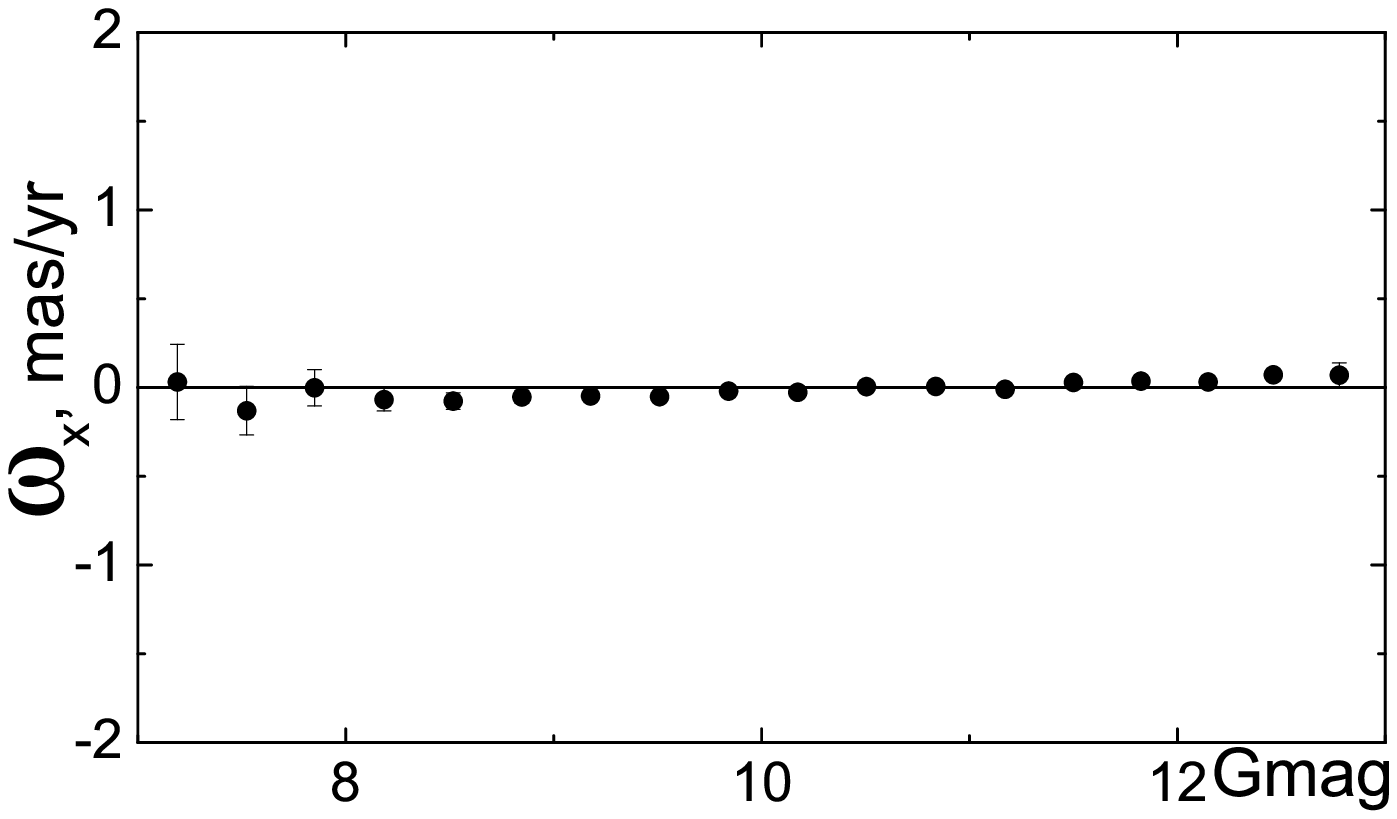}
\includegraphics[width = 58mm,]{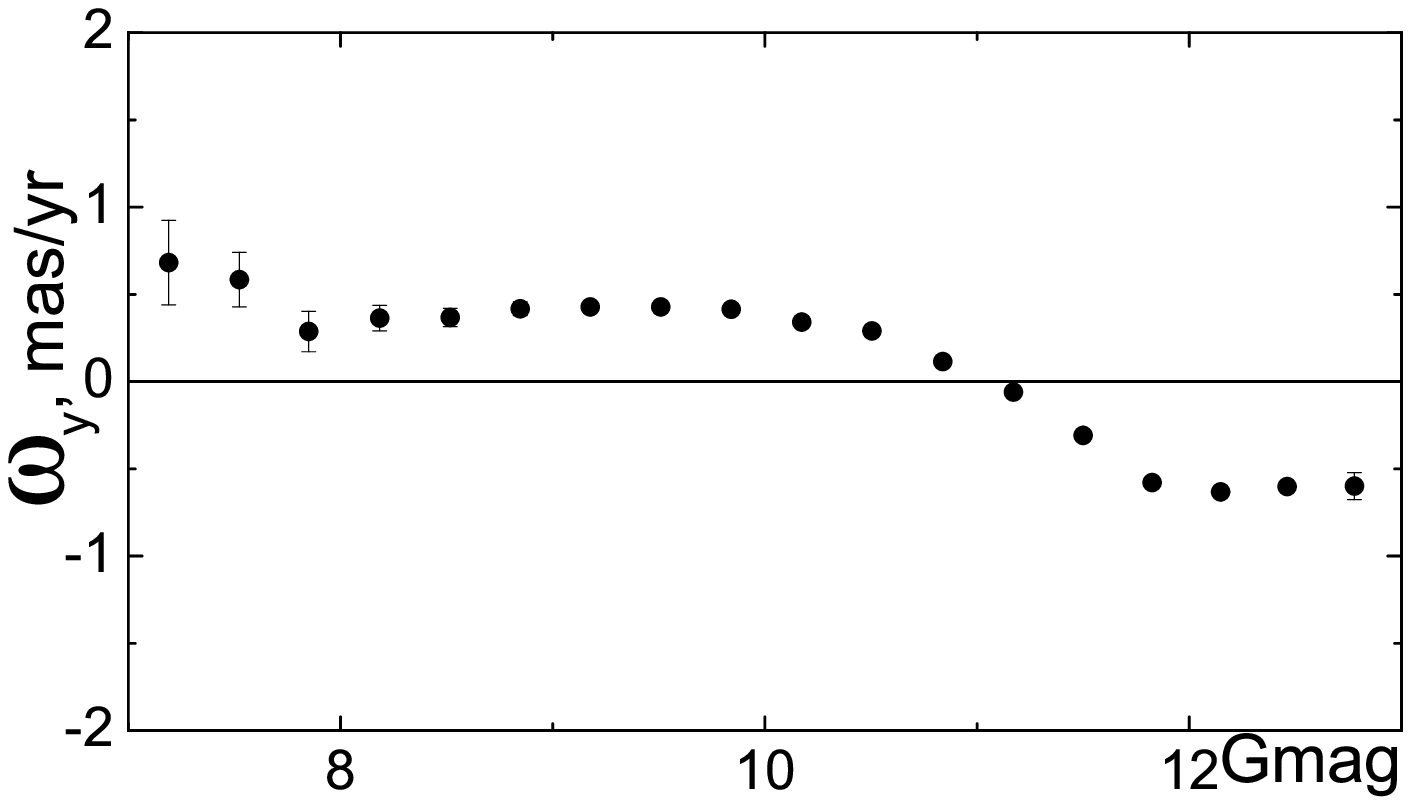}
\includegraphics[width = 58mm,]{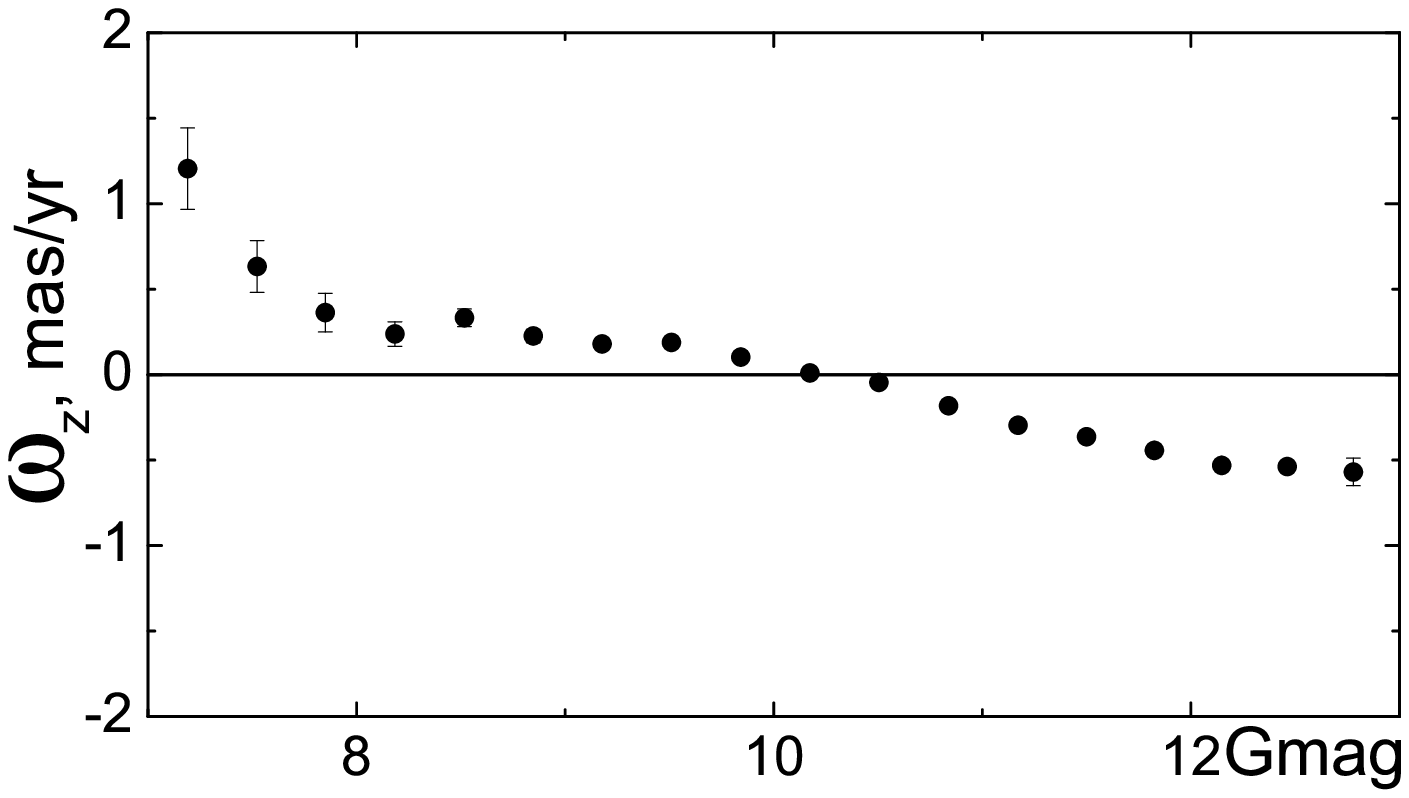}

\includegraphics[width = 58mm,]{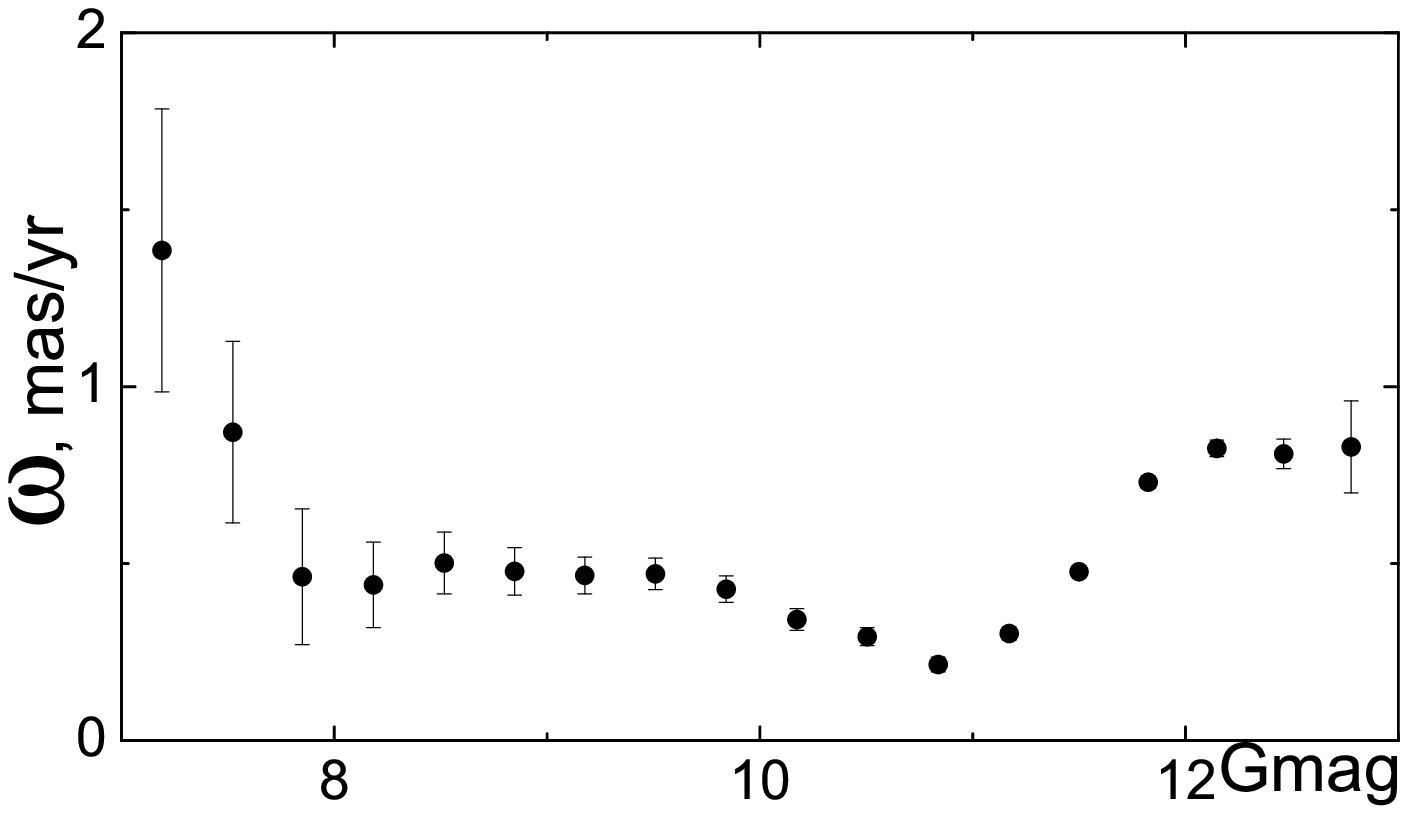}
\includegraphics[width = 58mm,]{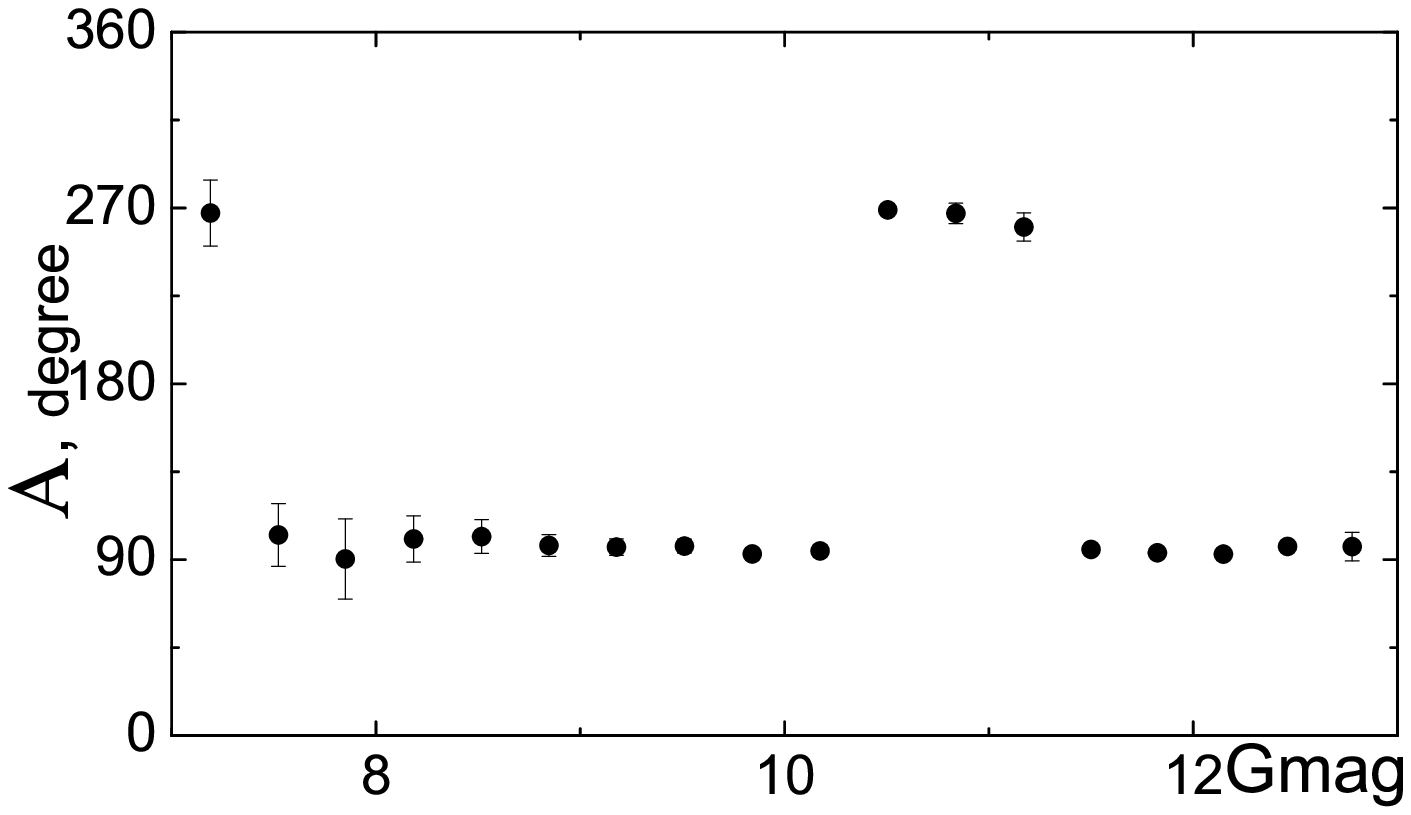}
\includegraphics[width = 58mm,]{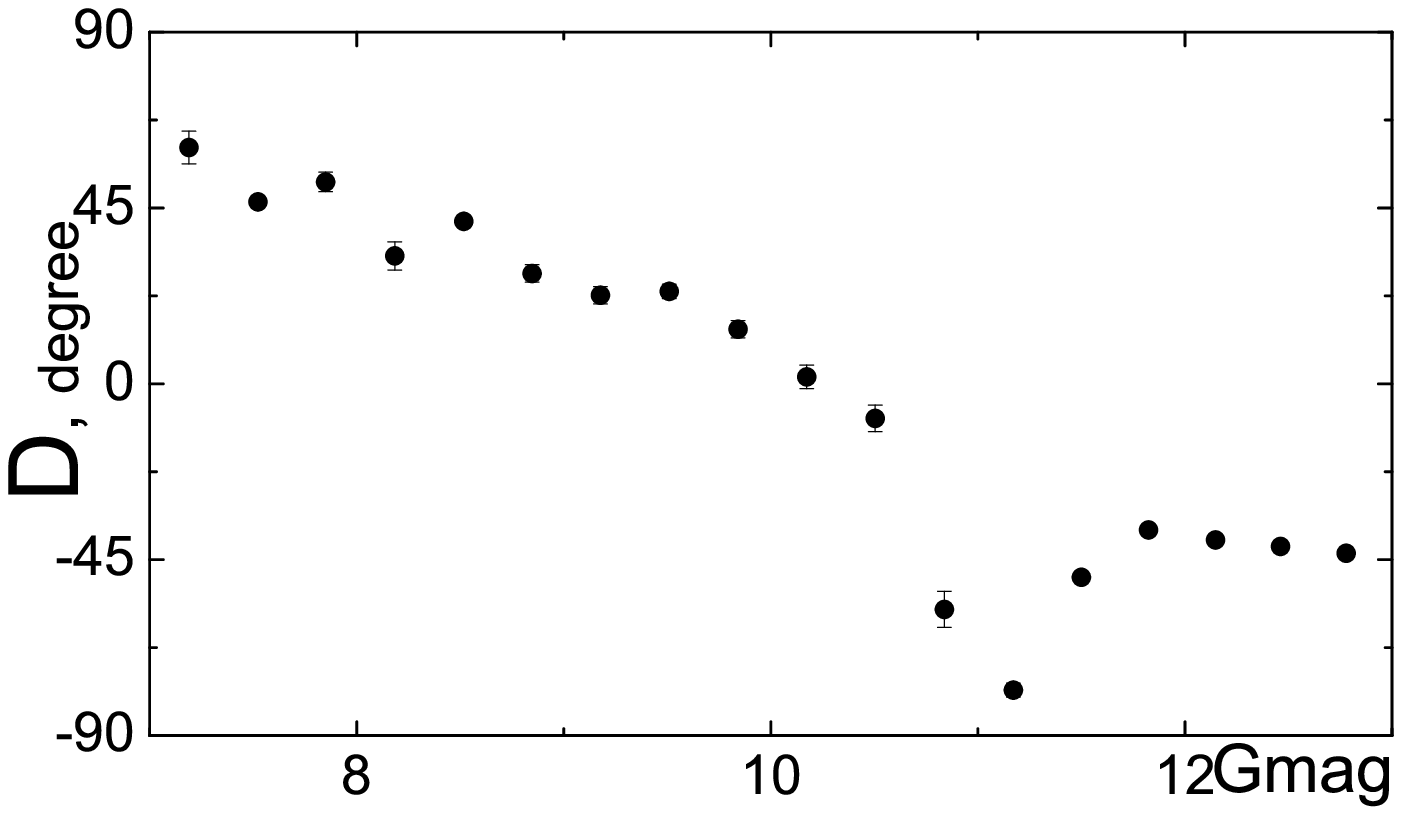}
\caption{Components of the mutual rotation vector between coordinate systems of the PMA and TGAS as a function of G magnitude (uppper panel). The modulus of mutual rotation vector and the values of pole's coordinates (bottom panel).}
\label{pto}
\end{figure*}

\begin{figure*}
\vspace*{0pt}
\includegraphics[width = 58mm,]{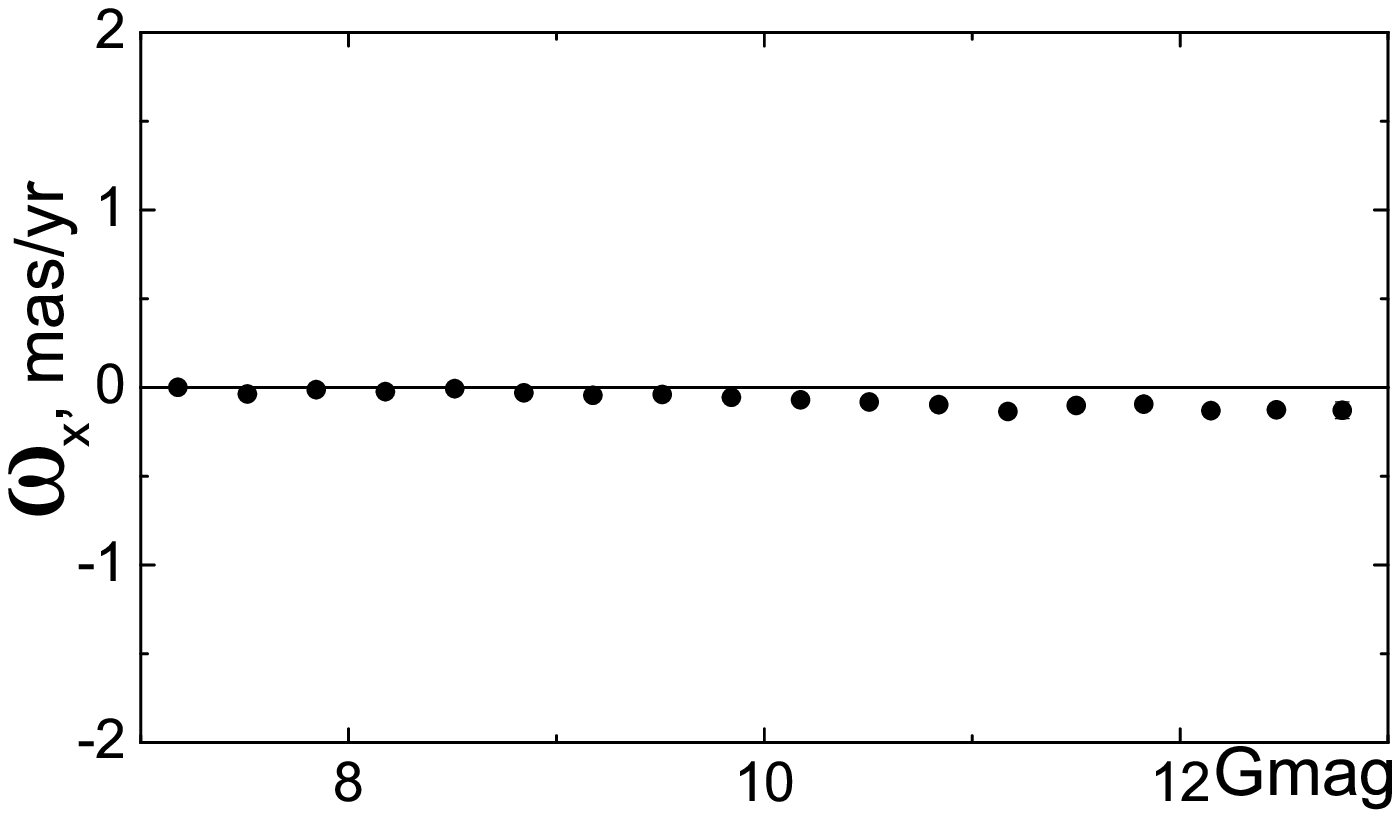}
\includegraphics[width = 58mm,]{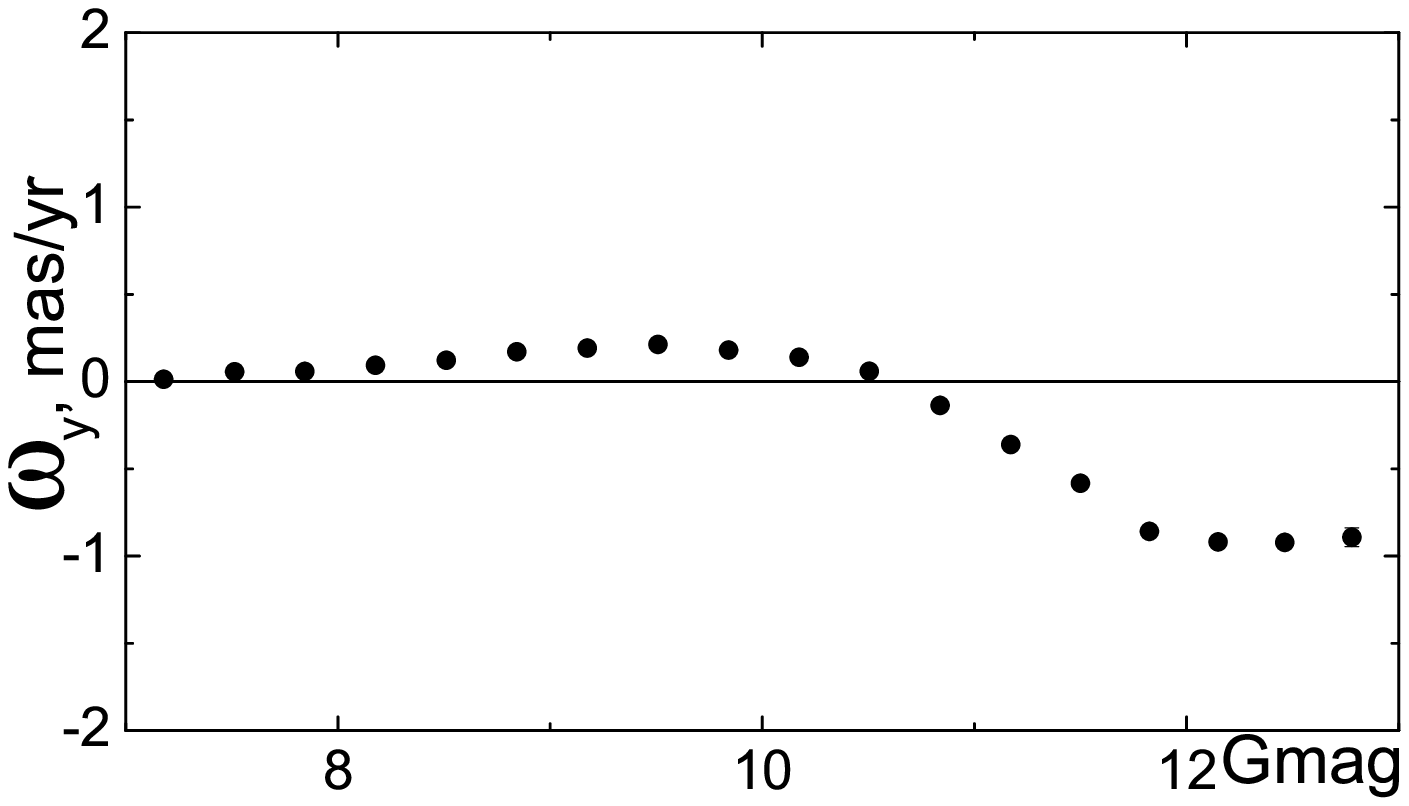}
\includegraphics[width = 58mm,]{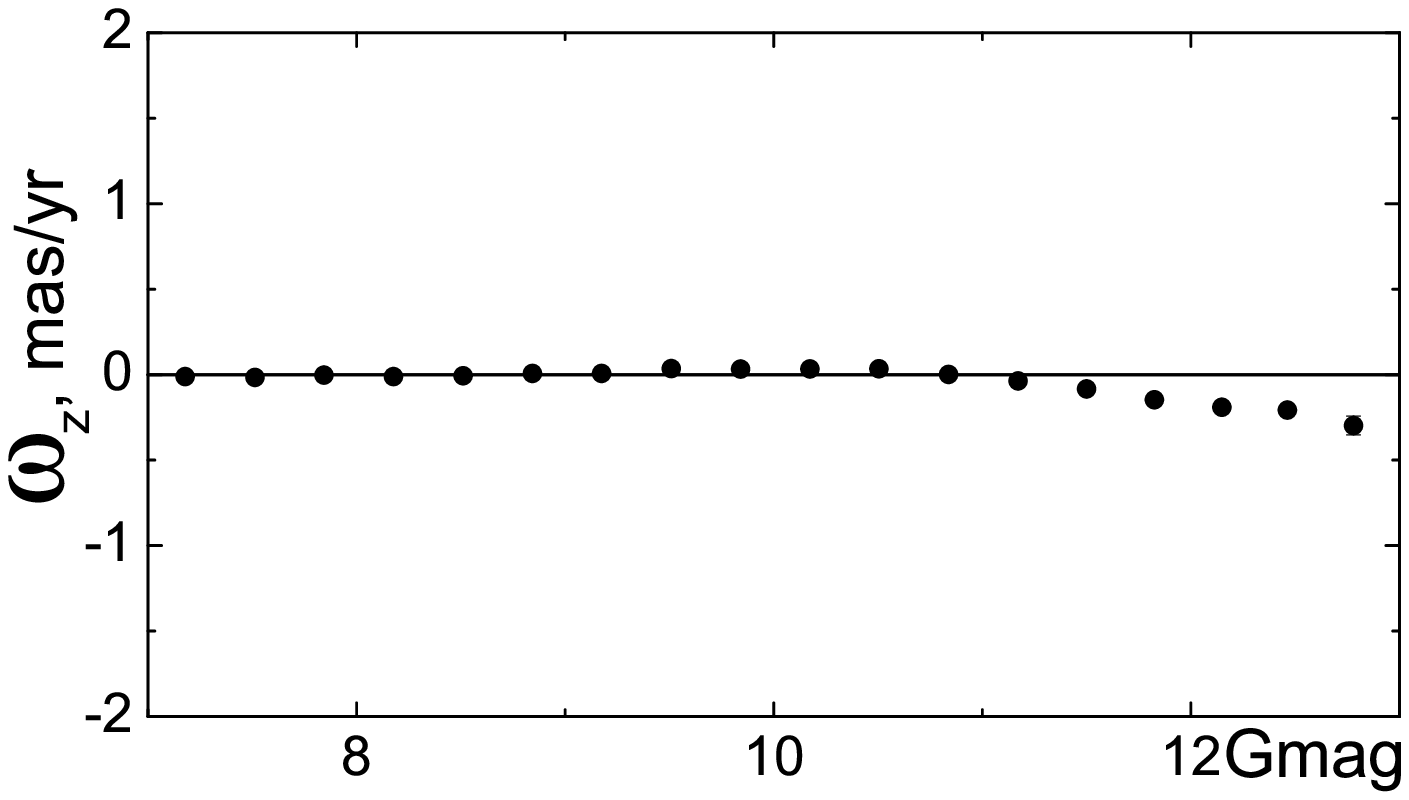}

\includegraphics[width = 58mm,]{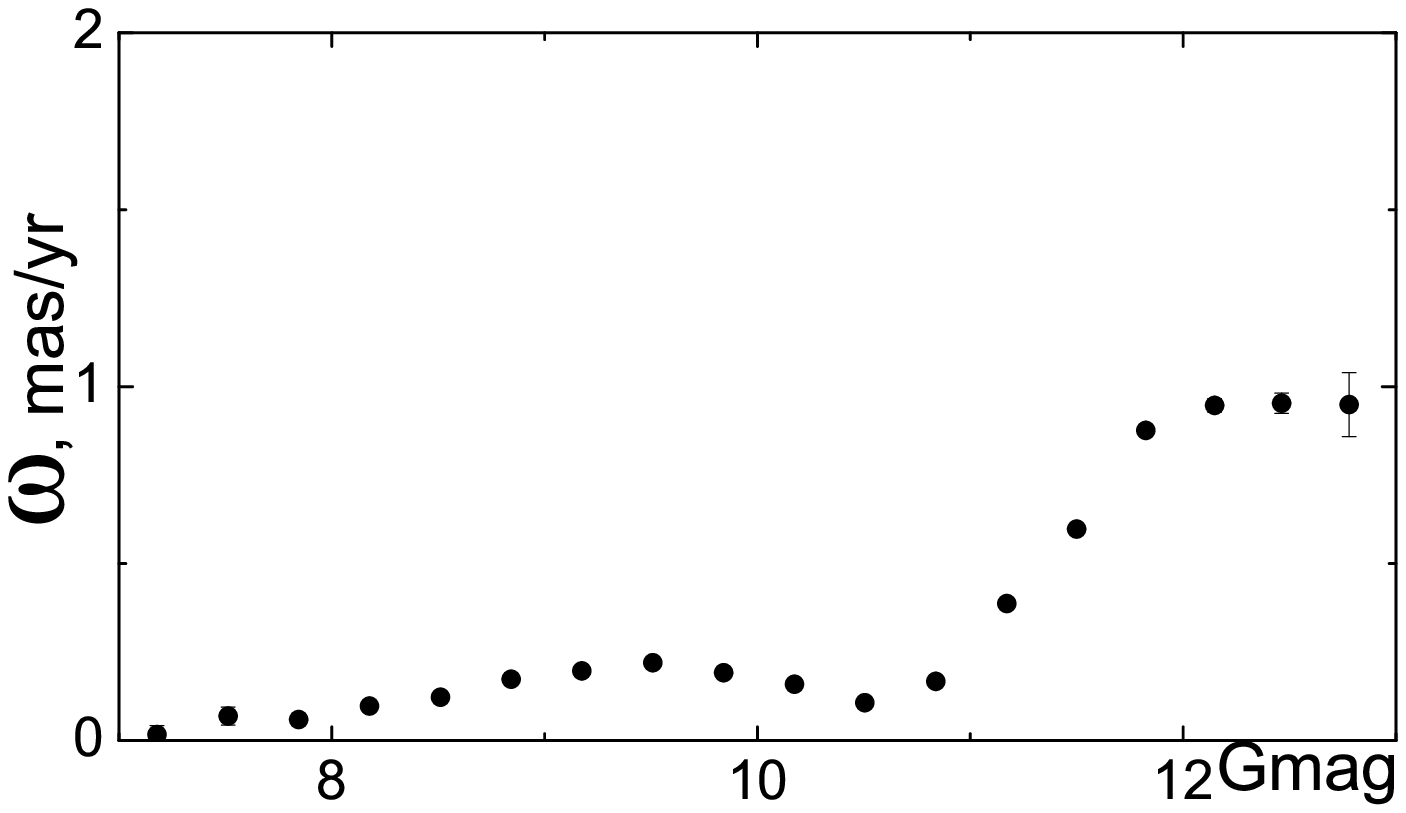}
\includegraphics[width = 58mm,]{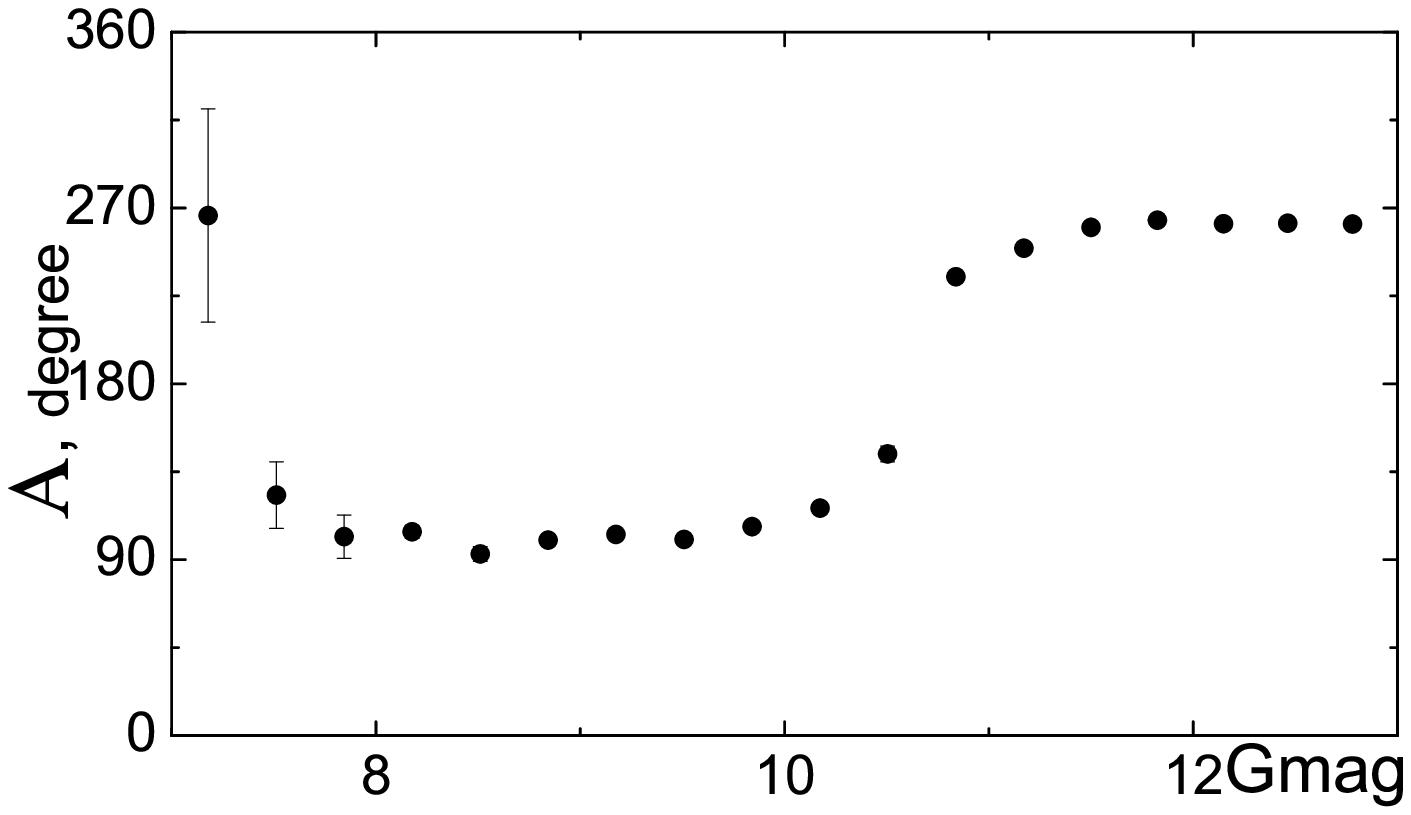}
\includegraphics[width = 58mm,]{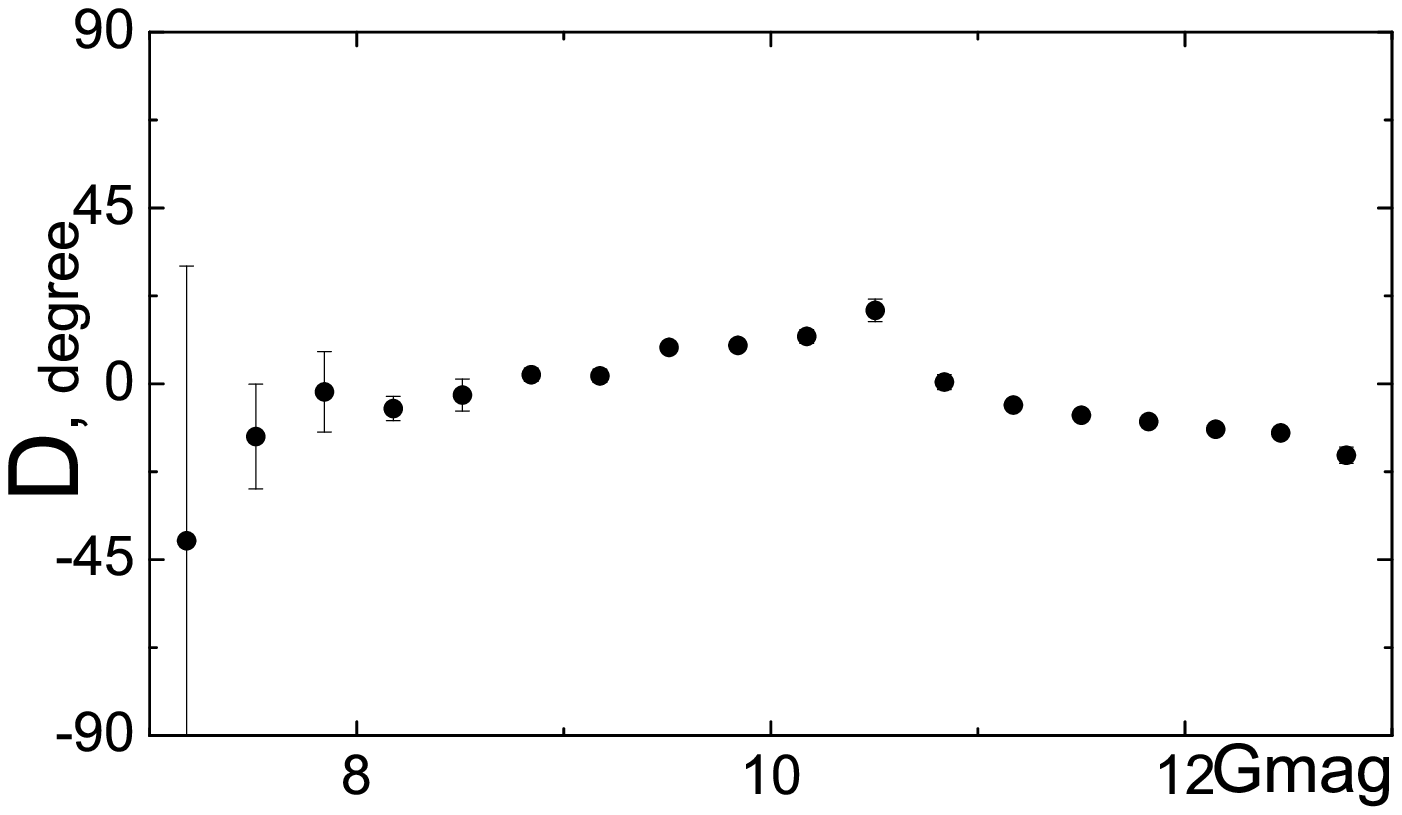}
\caption{Components of the mutual rotation  vector between coordinate systems of the HSOY and TGAS as a function of G magnitude (uppper panel). The modulus of mutual rotation vector and the values of pole's coordinates (bottom panel).}
\label{hto}
\end{figure*}

\begin{figure*}
\vspace*{0pt}
\includegraphics[width = 58mm,]{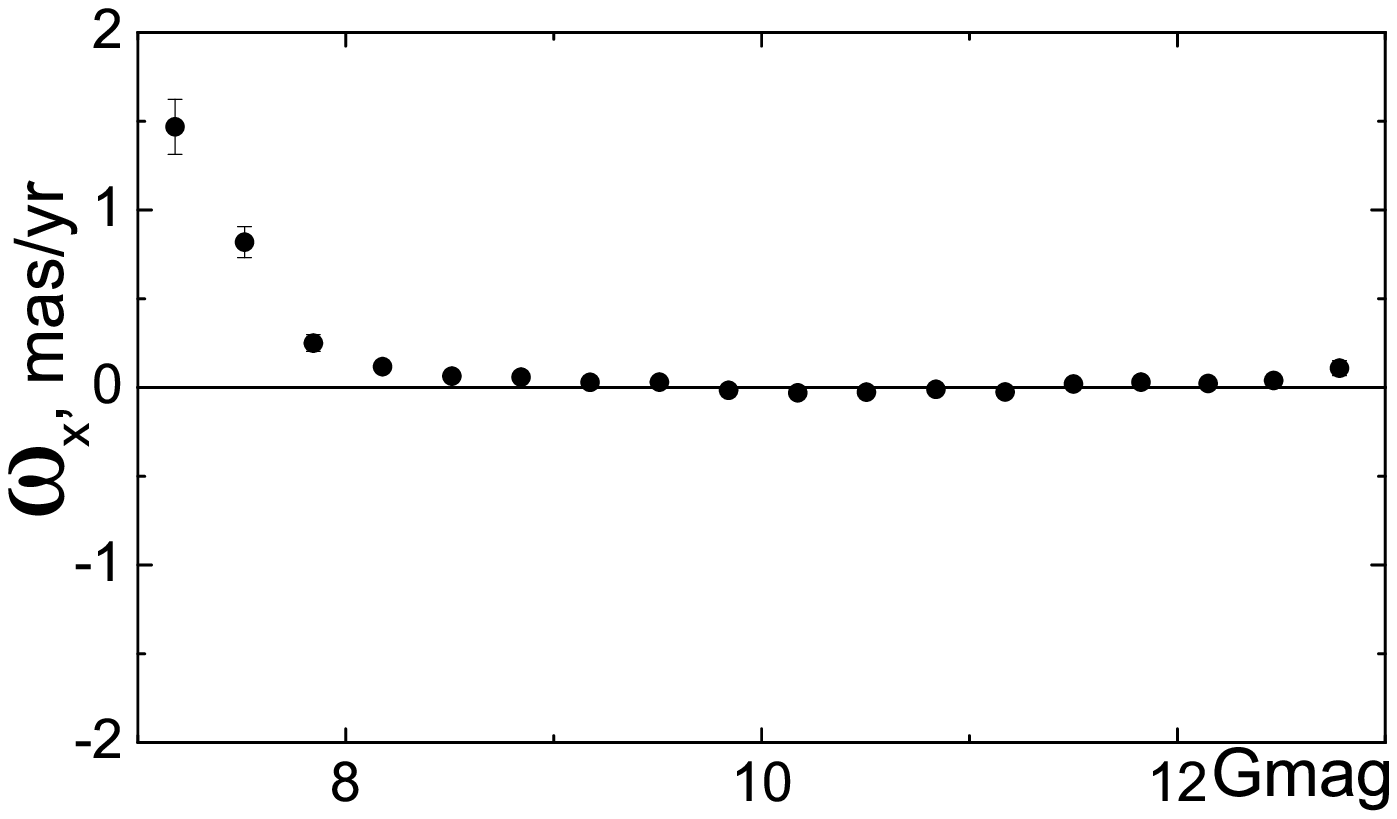}
\includegraphics[width = 58mm,]{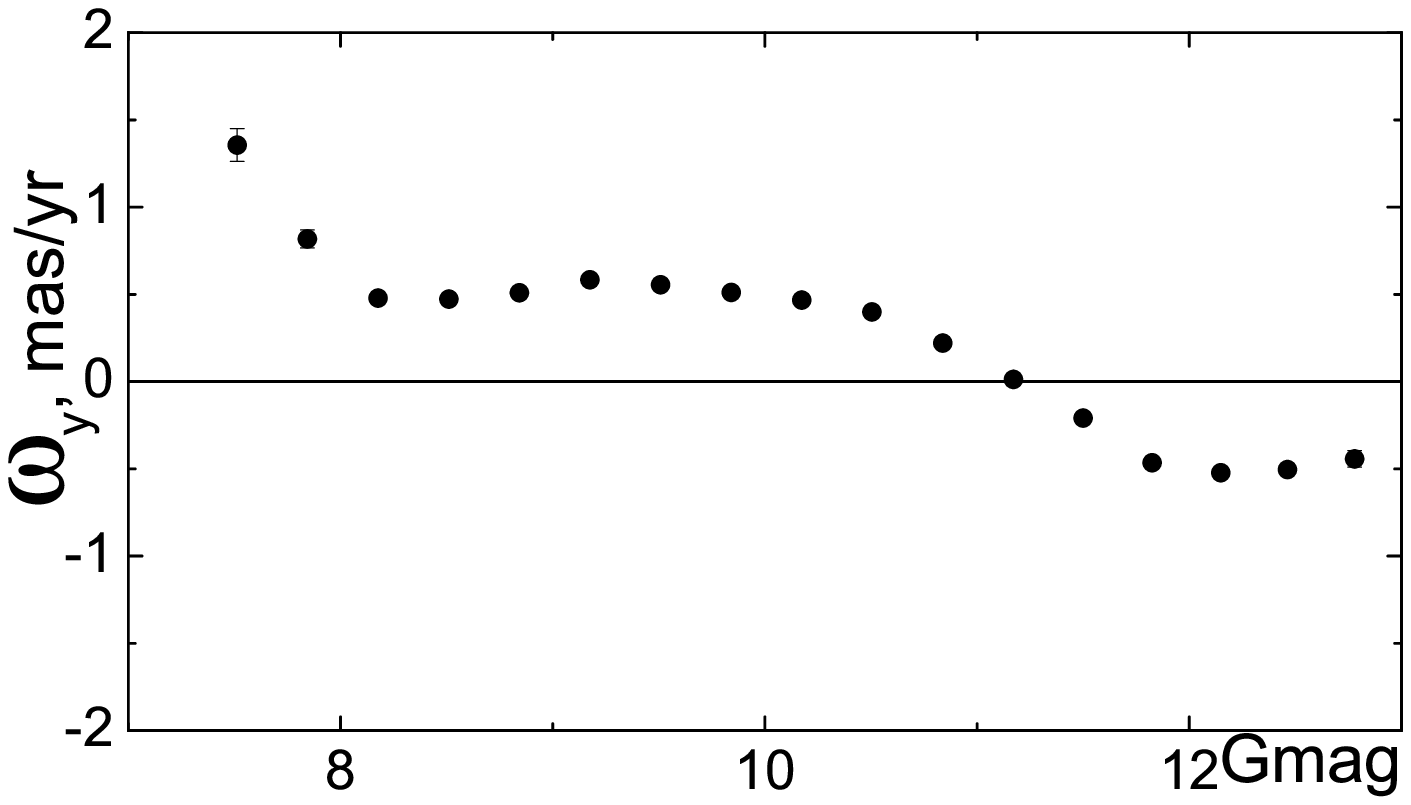}
\includegraphics[width = 58mm,]{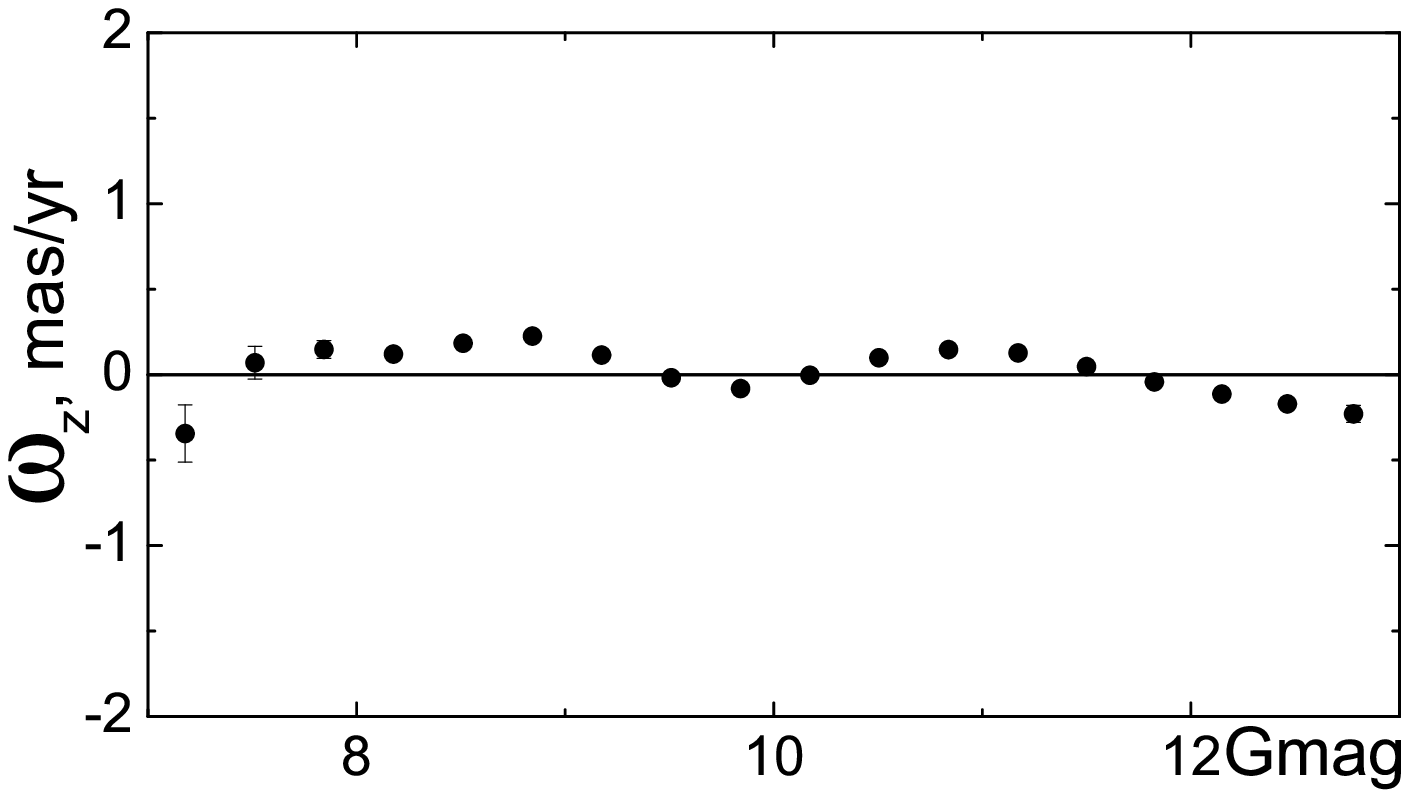}

\includegraphics[width = 58mm,]{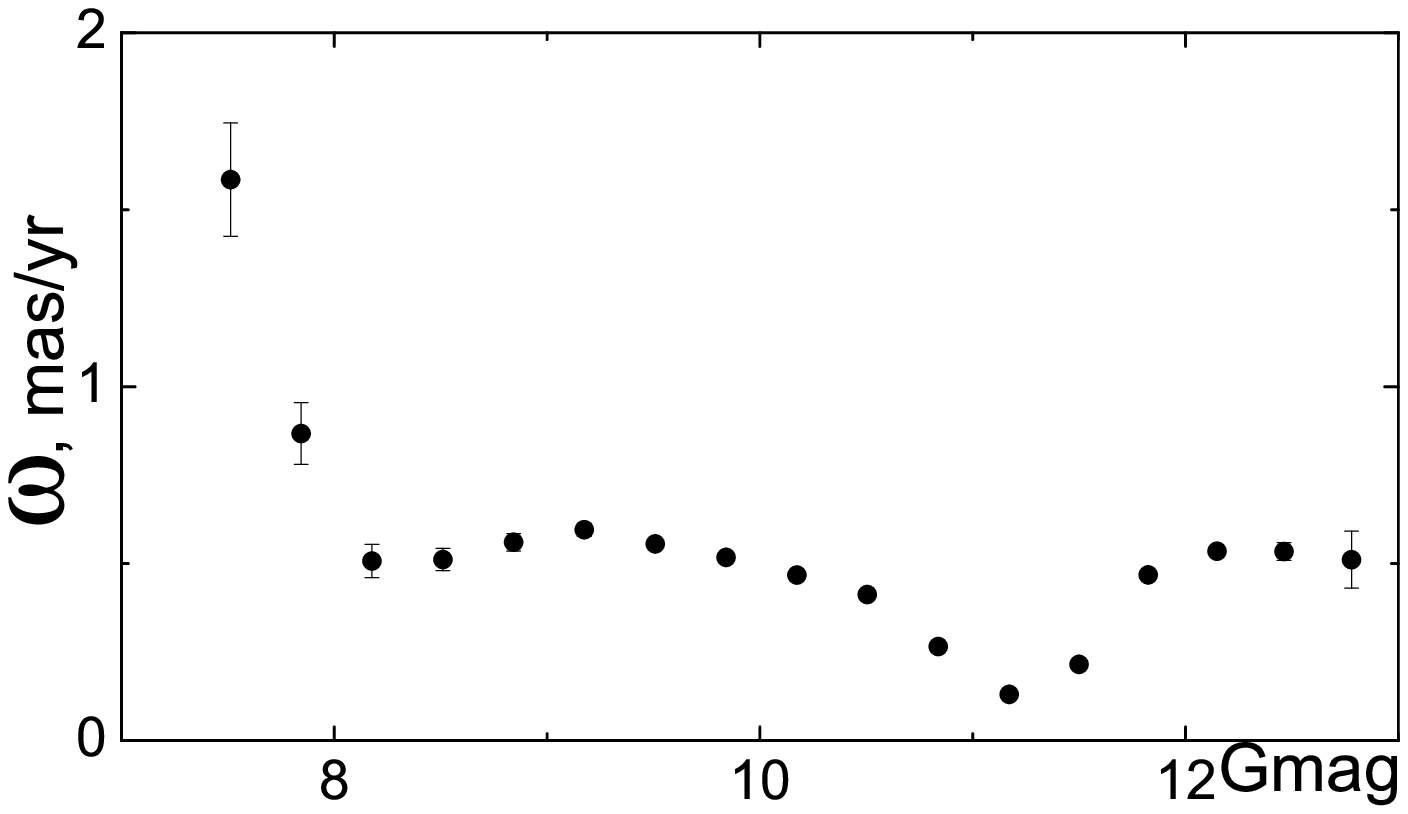}
\includegraphics[width = 58mm,]{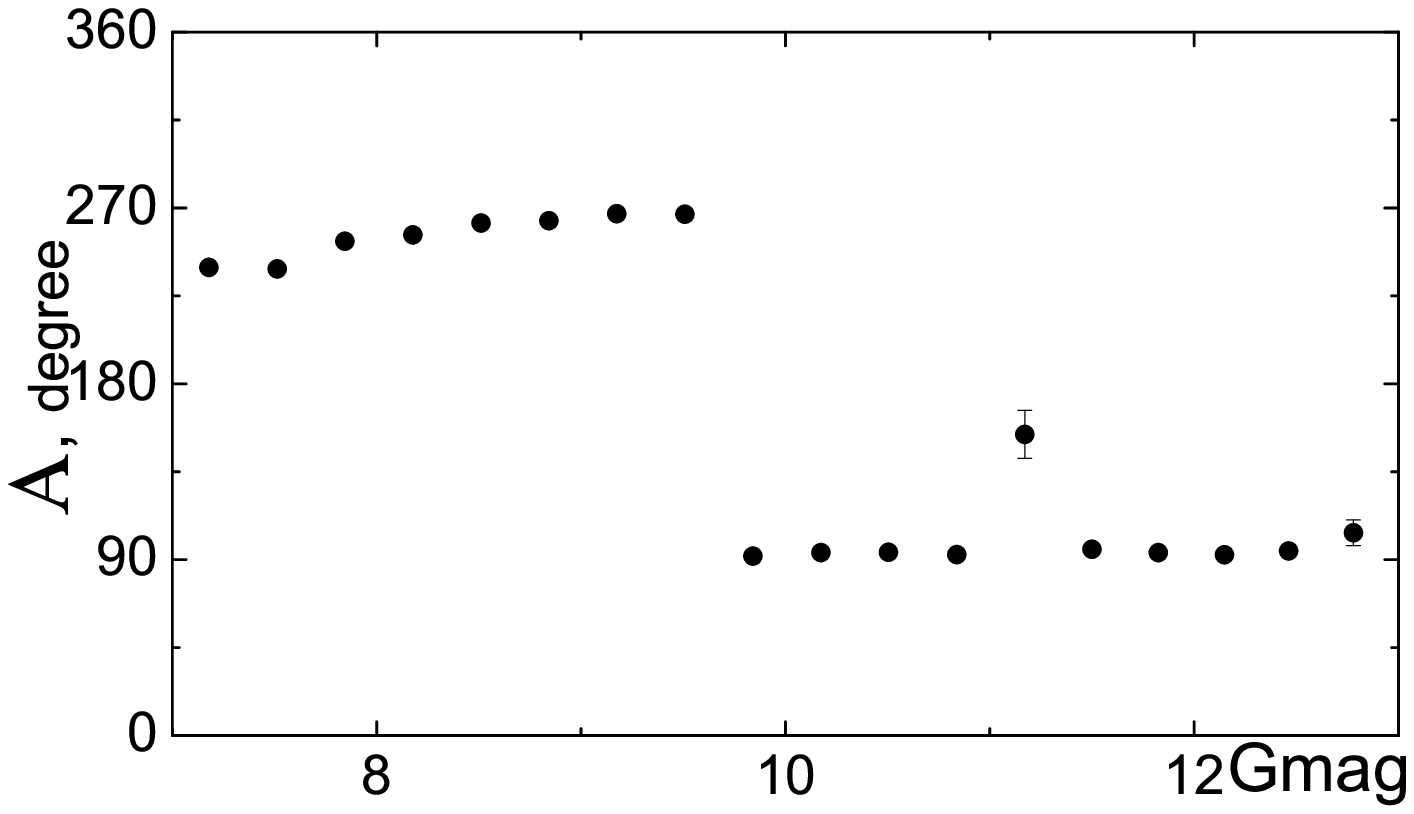}
\includegraphics[width = 58mm,]{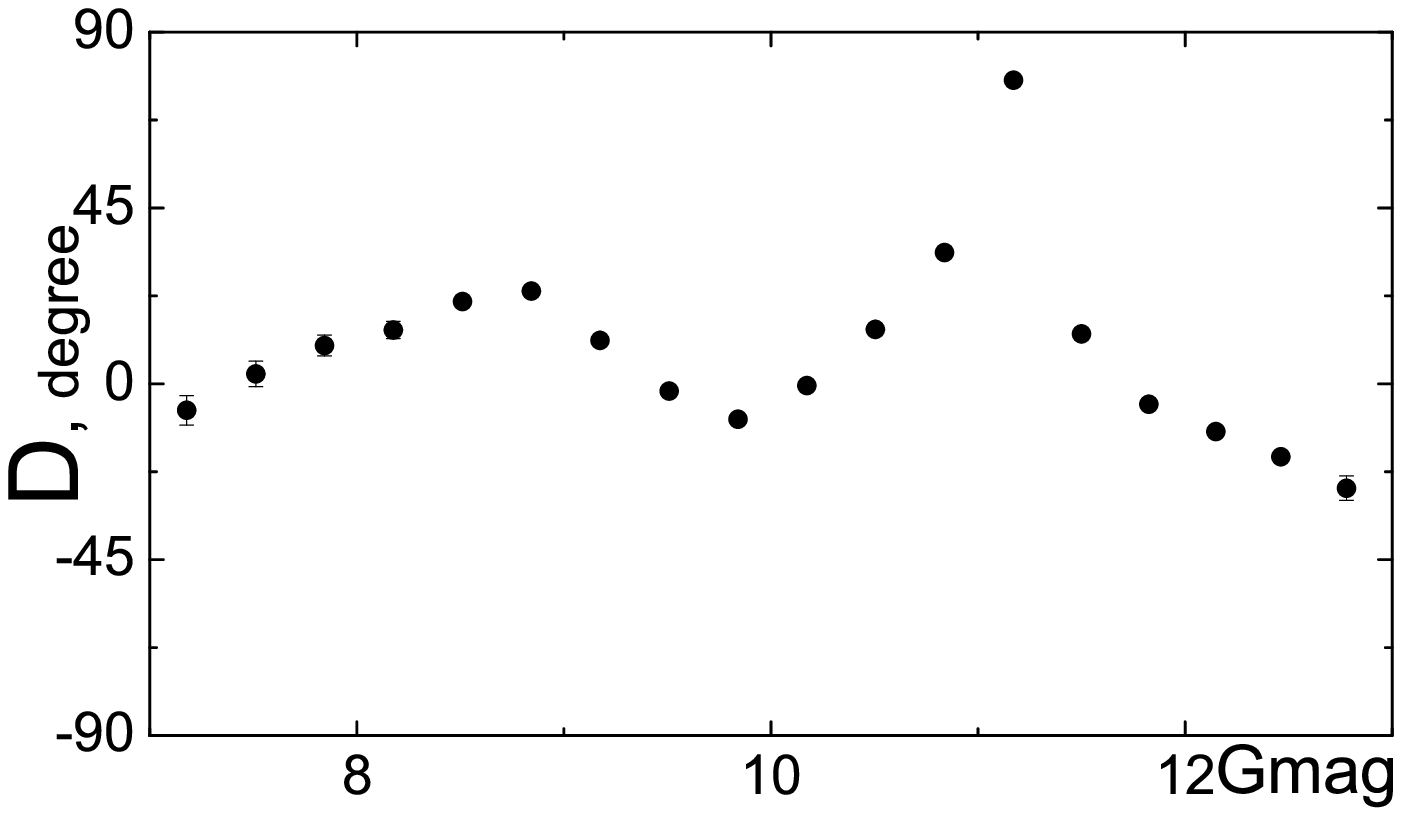}
\caption{ Components of the mutual rotation  vector between coordinate systems of the UCAC5 and TGAS as a function of G magnitude (uppper panel). The modulus of mutual rotation vector and the values of pole's coordinates (bottom panel).}
\label{uto}
\end{figure*}
 
As is known \citep {a5}, proper motions of the PMA catalogue are absolute, i.e. given in the extragalactic reference frame realized by positions of about 1.6 million galaxies from the Gaia DR1. Therefore we can consider them as independent relative to proper motions of the TGAS stars which, as is noted in \citep {l1}, are given in the Gaia DR1 reference frame which is in agreement at epoch J2015.0 with the ICRF2 quasars \citep {m5,f2} accurate to better than 0.1 mas and does not rotate relative to the ICRF2 within 0.03 mas/yr. In fact, difference between the specified frames is only that they are realized by various extragalactic sources. Because Gaia’s observational strategy was optimized for point sources \citep {p3}, differences between image profiles for quasars and galaxies in the Gaia DR1 have to be practically absent. It means that the coordinates of their centroides used to fix the TGAS and PMA reference frames can be slightly different due to precision error but systematically have to be close.
Comparison of stellar proper motions of the specified catalogues itself is of interest and involving new catalogues such as HSOY and UCAC5 using the TGAS catalogue as the reference one, extends capacity of the analysis and interpretation of the results derived.

First of all, we compare proper motions of the objects globally throughout the whole celestial sphere. In figures \ref{m_a_thup} dependencies of proper motion differences $\bigtriangleup\mu_{\alpha} cos \delta$ and $\bigtriangleup\mu_{\delta}$  depending on magnitude are presented.

For more detailed analysis we use the solid-body mutual rotation model of two coordinate systems \citep {l4} set by catalogues used. In approximation of infinitesimal orientation angles and assuming that the orientation parameters are functions of time, equations for differences between stellar proper motions along the right accention and declination are generally as follow 

\begin{equation}
{(\mu_{\alpha}^{CAT}-\mu_{\alpha}^{TGAS})cos\delta = \omega_x cos\alpha sin\delta + \omega_y sin\alpha sin\delta - \omega_z cos \delta} 
\end{equation} 
\begin{equation}
{(\mu_{\delta}^{CAT}-\mu_{\delta}^{TGAS}) = - \omega_x sin \alpha + \omega_y cos \delta}
\end{equation}

where $\omega_{x}$, $\omega_{y}$, $\omega_{z}$  are components of the angular velocity vector of the rotation Cartesian coordinate system set by the CAT catalogue relative to the TGAS catalogue coordinate system.

In figure \ref{pto} (uppper panel) components of the rotation vector derived for the whole celestial sphere by joint solutions of the equation system (1) and (2) are presented.
From these figures it is clearly seen that in bright part of the magnitude range $7.5^{m}<G<9^{m}$  all the three components of the rotation vector practically do not depend on magnitude and on module do not exceed 0.5 mas/yr, although slightly differ by quantity. At the same time, beginning from $9^{m}$ behaviour of components clearly shows variations depending on magnitude, while beginning from $11.5^{m}$ this dependencies are virtually vanished. Especially strange is the dependence of $\omega_{y}$, demonstrating a sign change near 11 magnitude.

\subsection{Investigation of proper motions from the UCAC5 and HSOY}

At first, it should be noted that if the components of the rotation vector depend on some parameters, for example, on the magnitude of the star, then the rotation will no longer be solid, since stars of different magnitude will cause different rotational velocities of the coordinate systems.
Therefore we assumed that the detected dependencies of $\omega_{y}$ and $\omega_{z}$ on magnitude were caused by the magnitude equation in proper motions of the PMA. To check this assumption we also have analyzed vectors of mutual rotation between systems of the UCAC5 and TGAS, HSOY and TGAS. The corresponding plots for components of the rotation vectors between the systems are given in figures \ref{hto}, \ref{uto}. 
As it can be seen from these plots, the $\omega_{x}$ component in both cases does not depend on magnitude. The $\omega_{z}$ component when comparing with the UCAC5 catalogue, though changes in a complicated manner in the $7.5 < m < 13$ range, can be considered as a constant value within $\pm$ 0.25 mas/yr. When comparing with the HSOY catalogue $\omega_{z}$ practically does not depend on magnitude. Behaviour of the $\omega_{y}$ component in both cases virtually coincides with its behaviour in figures \ref{pto}. This indicates similarity of the magnitude equations in proper motions of the ground-based PMA, UCAC5 and HSOY catalogues. At first sight, it is not surprising that the system of «space» proper motions (GAIA) differs from the «ground-based» one, for example, due to magnitude equation errors, usually present in ground-based observations. However taking into account that the stellar proper motions of the PMA, UCAC5 and HSOY are independent because of the use of different methods of their derivation and 
different initial data, we suppose that simultaneous existence of identical magnitude equations in three different catalogues is unlikely. Therefore, we have made the following assumption: dependence of the $\omega_{y}$ component on magnitude is caused by some systematics containing in proper motions of the TGAS stars. The reason for such an assumption was comparison of positions of Tycho-2 stars from the TGAS translated by the TGAS proper motions from epoch J2015.0 to the effective epoch $(T_{\alpha}+T_{\delta})/2$ near 1991 year with the positions taken from the Tycho-2 catalogue at epoch $(T_{\alpha}+T_{\delta})/2$. It turned out that the positional differences reach 5-7 mas. The same procedure applied to Hipparcos stars has shown that positional differences are less than 0.01 mas.

\section{Analysis of differences of the original and classical TGAS proper motions.}

As it is known  \citep {l1}, to derive positions, parallaxes and proper motions for about 2 million TGAS sources positional information from Hipparcos at the epoch J1991.25 and Tycho-2 at the observational epoch $(T_{\alpha}+T_{\delta})/2$ near 1991 year was used. Wherein in the paper's text is stressed that the proper motions derived with the use of the adaptive astrometric global iterative solution (AGIS) \citep{m4} are mean motions of stars between the two epochs J1991.25 and J2015.0 rather than their instantaneous proper motions in 2015 year. Median uncertainty of individual proper motions is 0.07 mas/yr for the Hipparcos stars and 1.4 mas/yr for non-Hipparcos Tycho-2 stars.

\begin{figure*}
\vspace*{0pt}
\includegraphics[width = 58mm,]{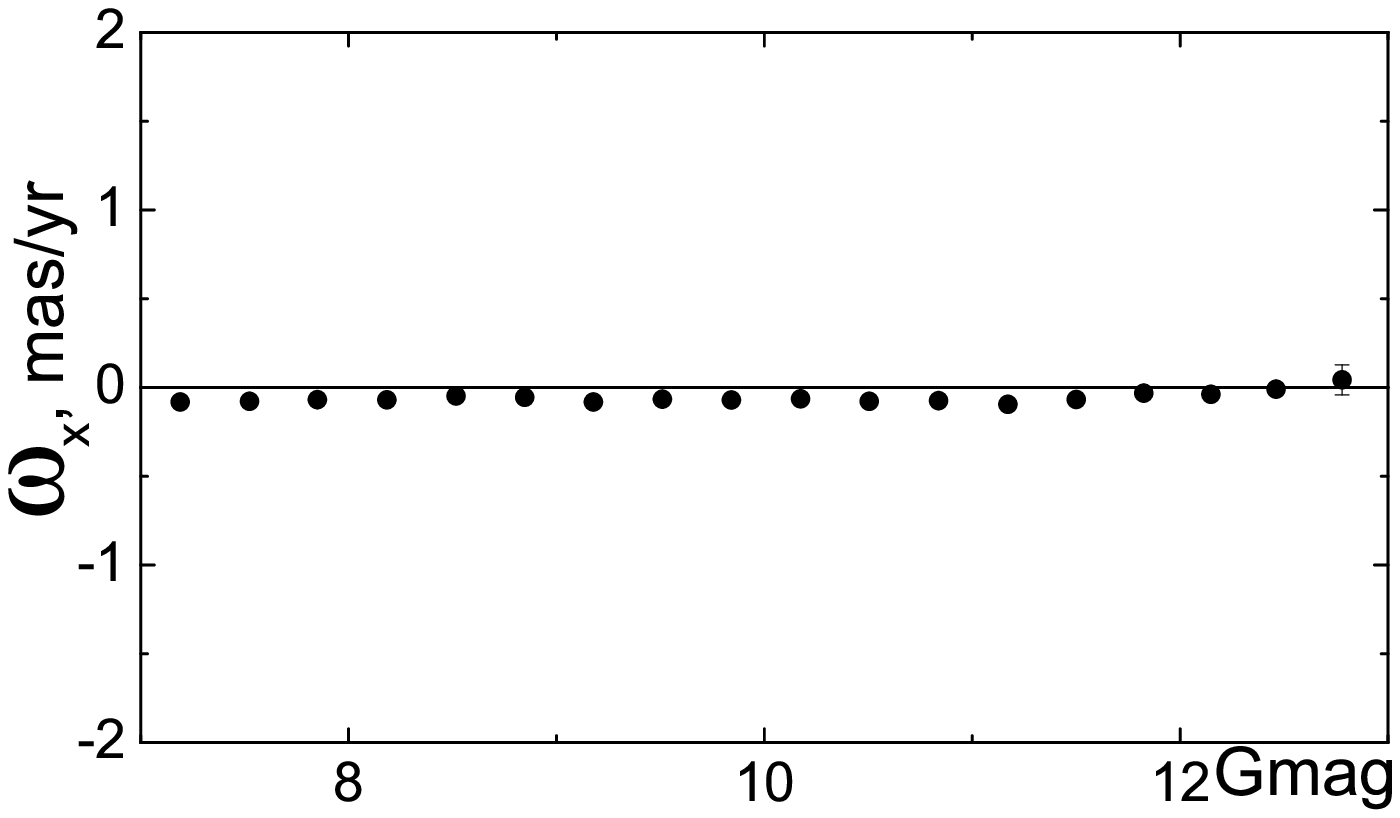}
\includegraphics[width = 58mm,]{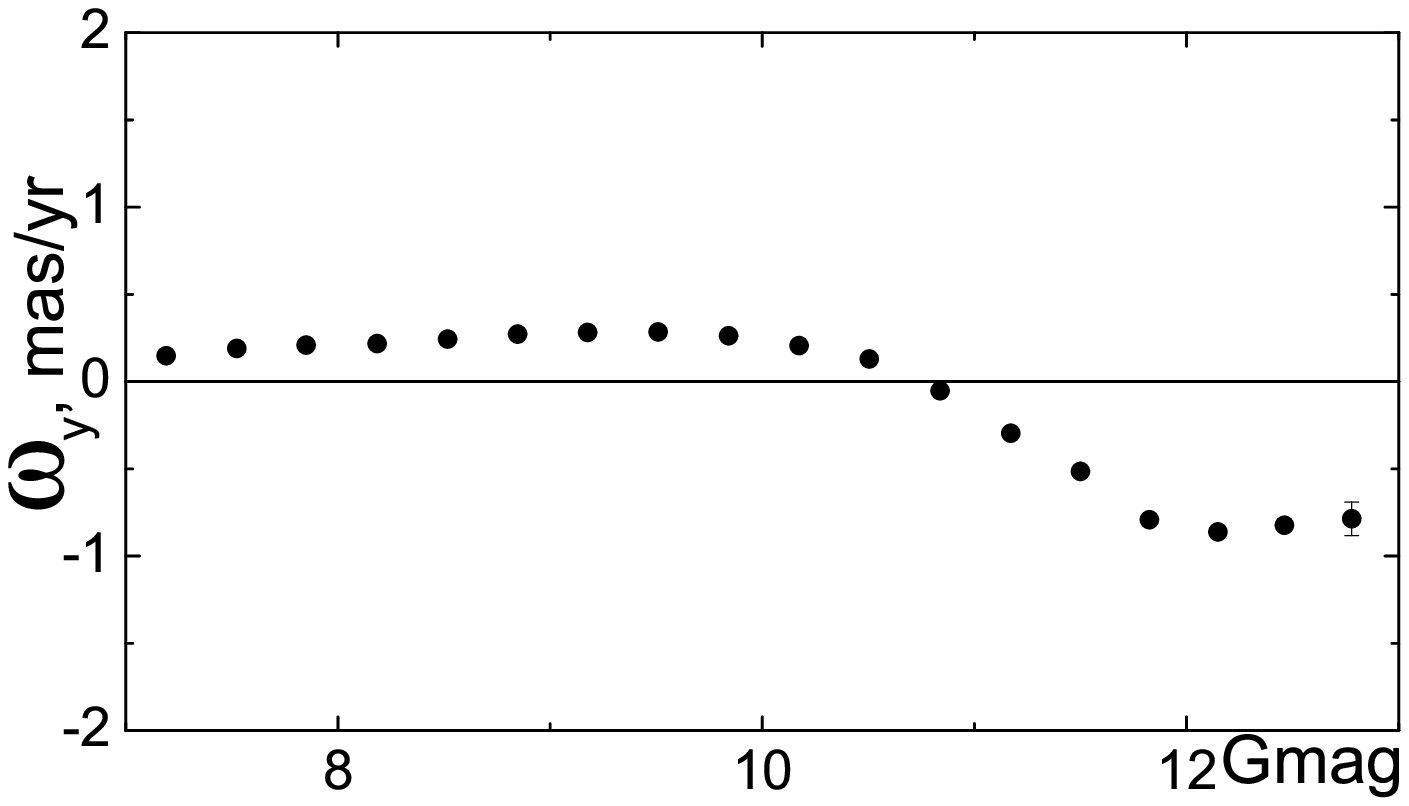}
\includegraphics[width = 58mm,]{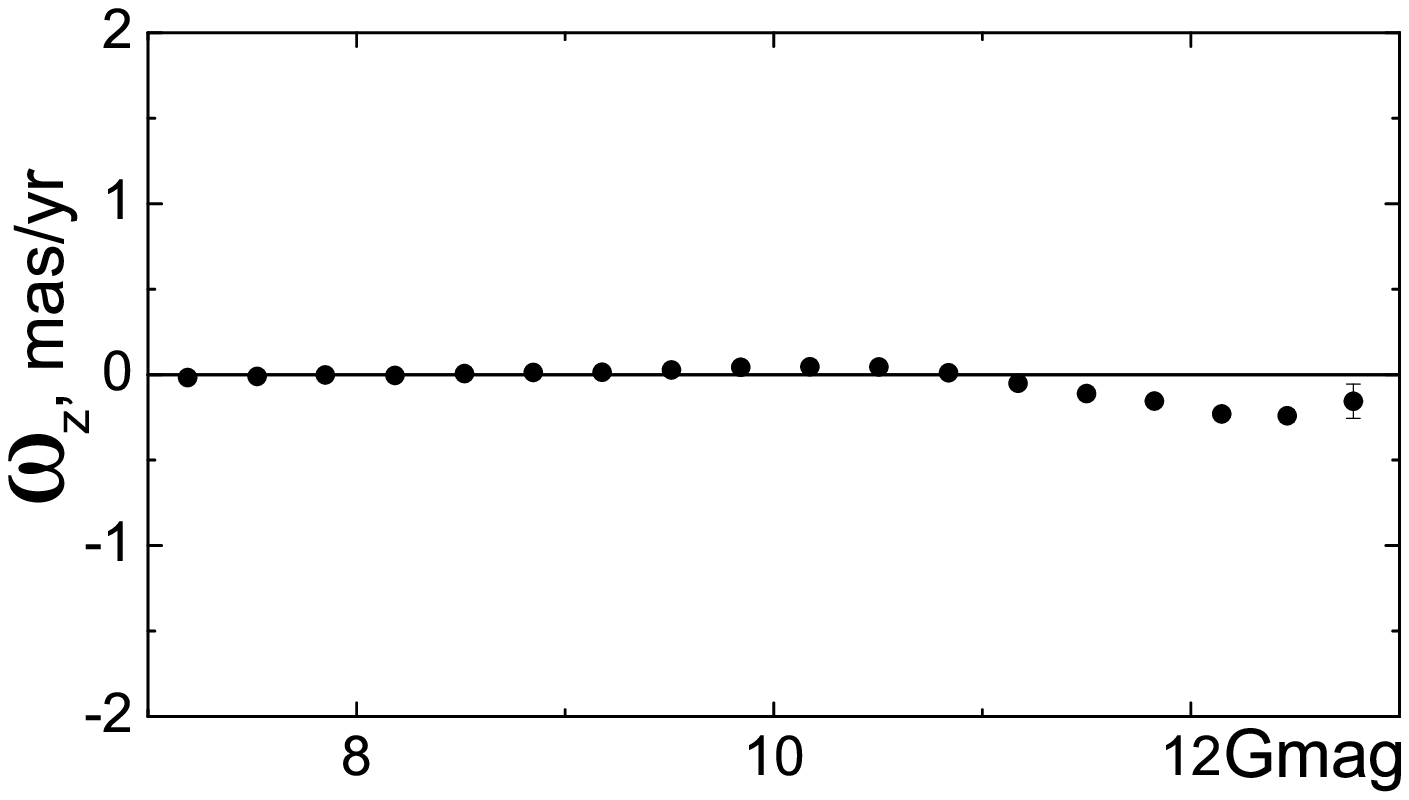}

\includegraphics[width = 58mm,]{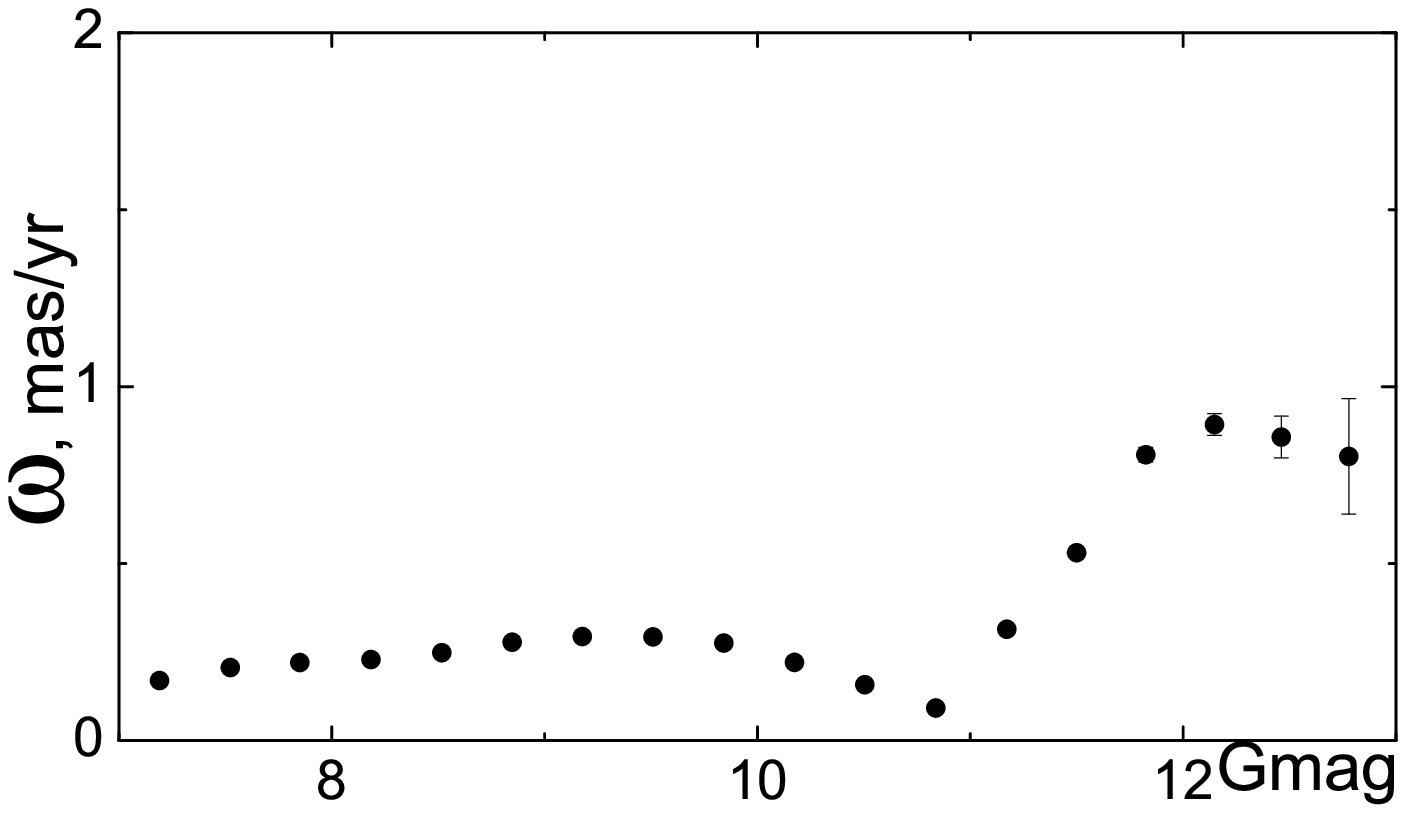}
\includegraphics[width = 58mm,]{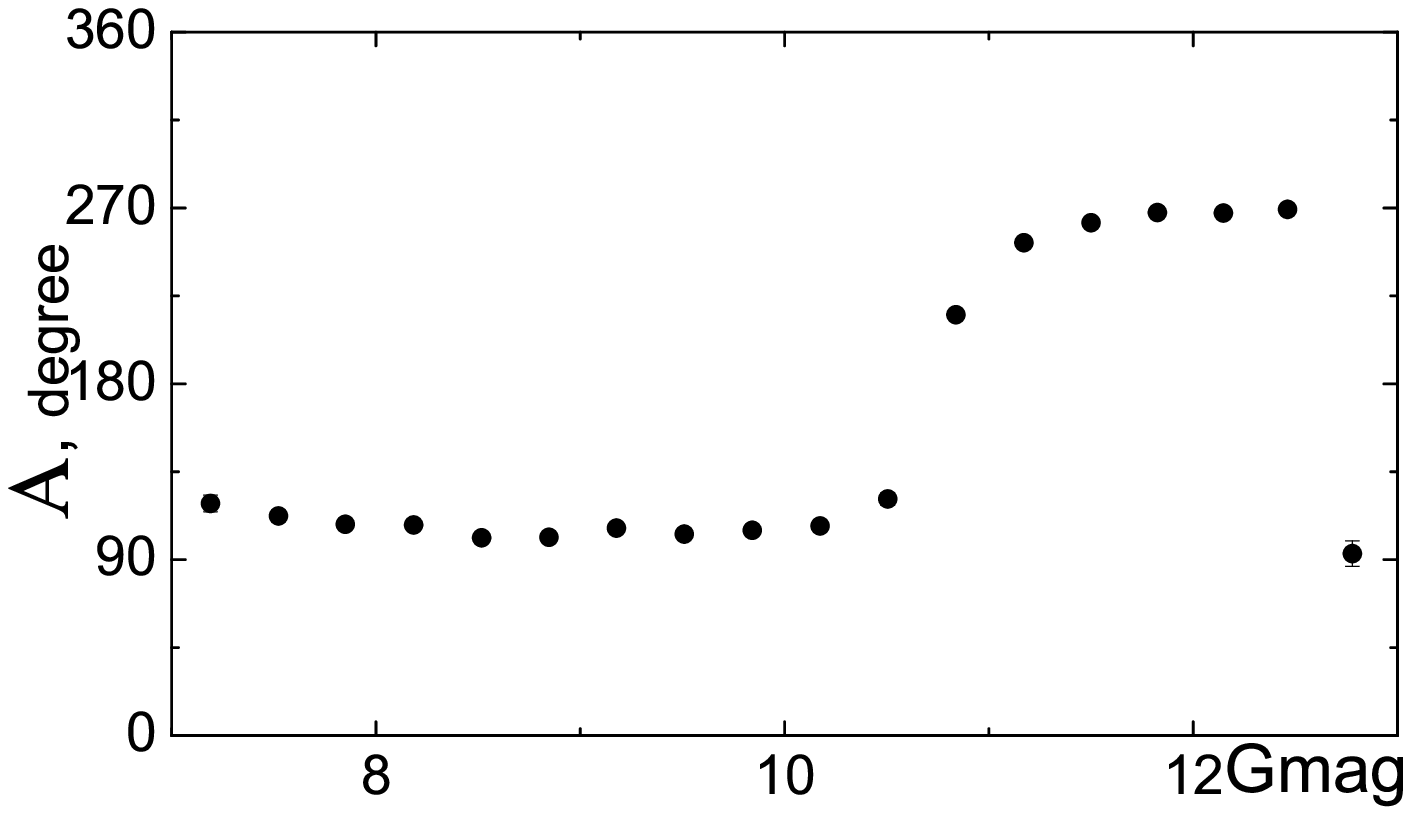}
\includegraphics[width = 58mm,]{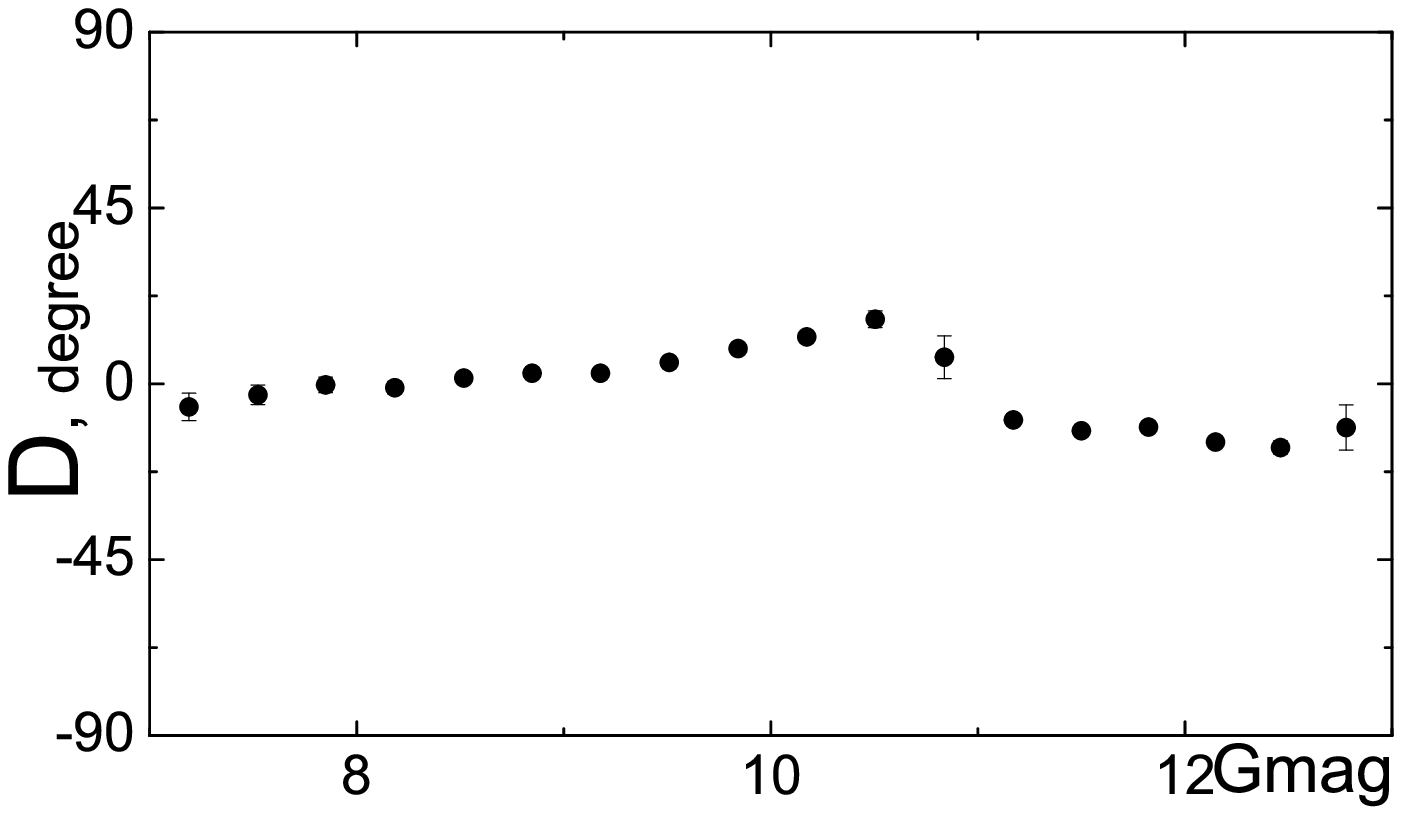}
\caption{Components of the mutual rotation  vector between coordinate systems of the TGAS$_{origin}$ (with proper motions of Tycho-2 stars derived in AGIS) and TGAS$_{classic}$ (with proper motions of Tycho-2 stars calculated by the classical method) as a function of G magnitude (uppper panel). The modulus of mutual rotation  vector and the values of pole's coordinates (bottom panel).} 
\label{totc}
\end{figure*}

\begin{figure*}
\vspace*{0pt}
\includegraphics[width = 58mm,]{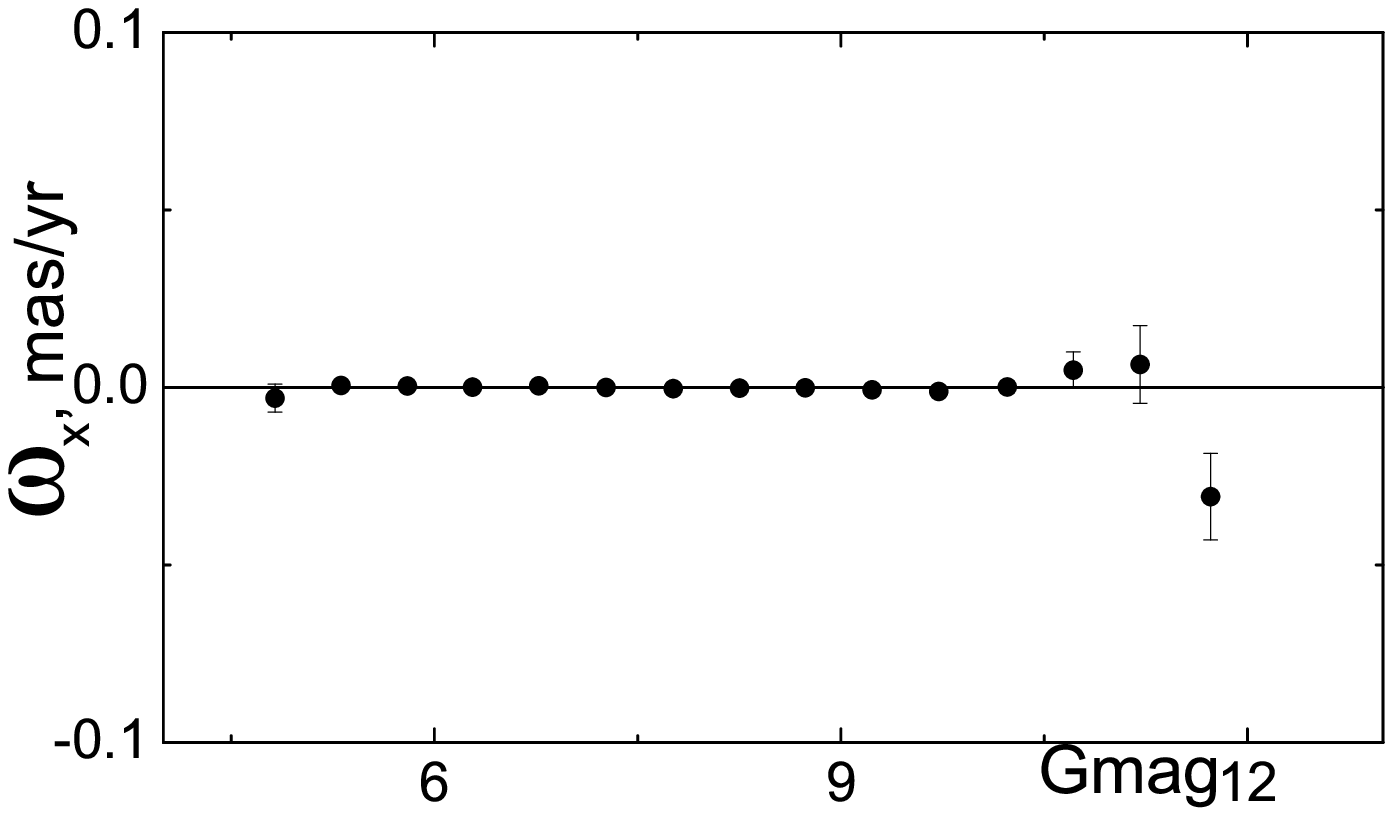}
\includegraphics[width = 58mm,]{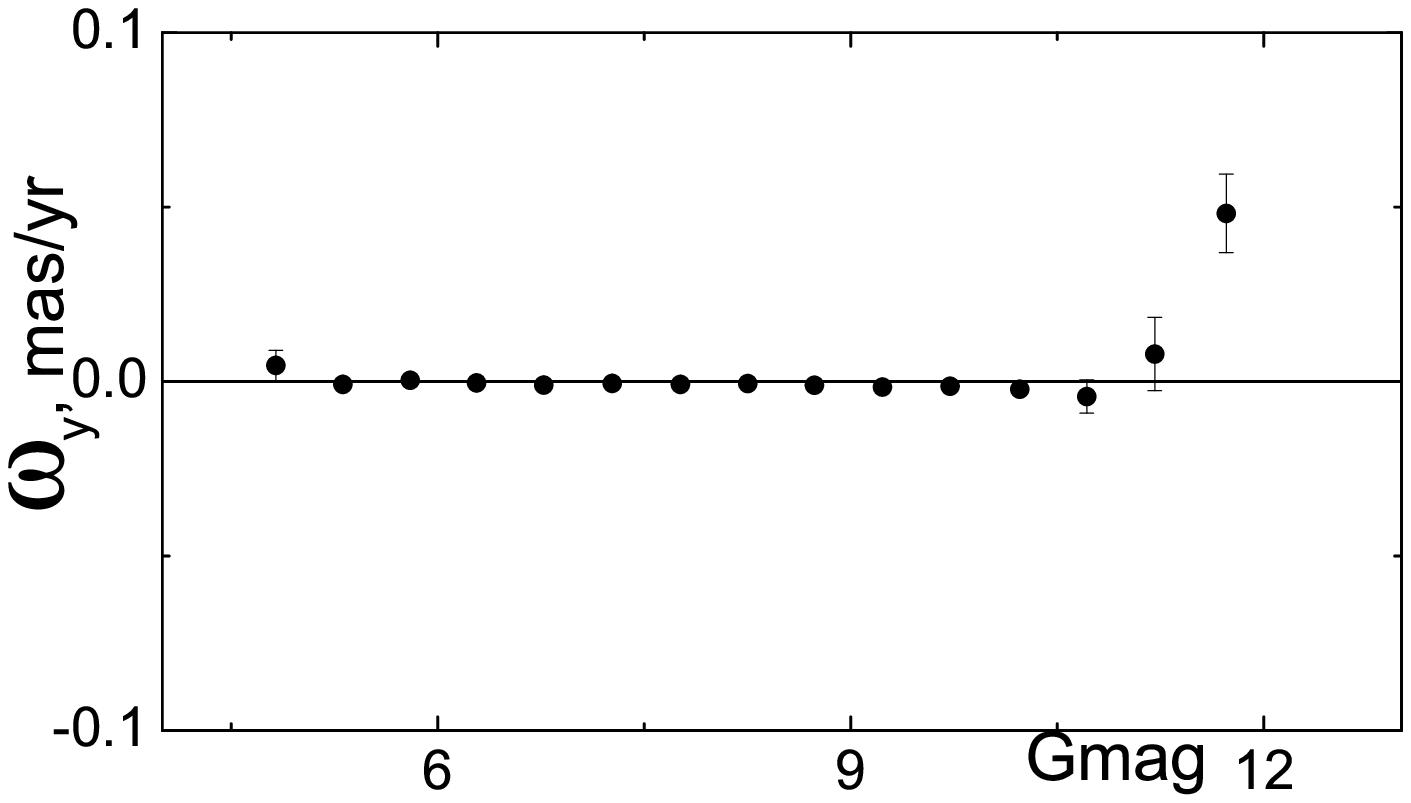}
\includegraphics[width = 58mm,]{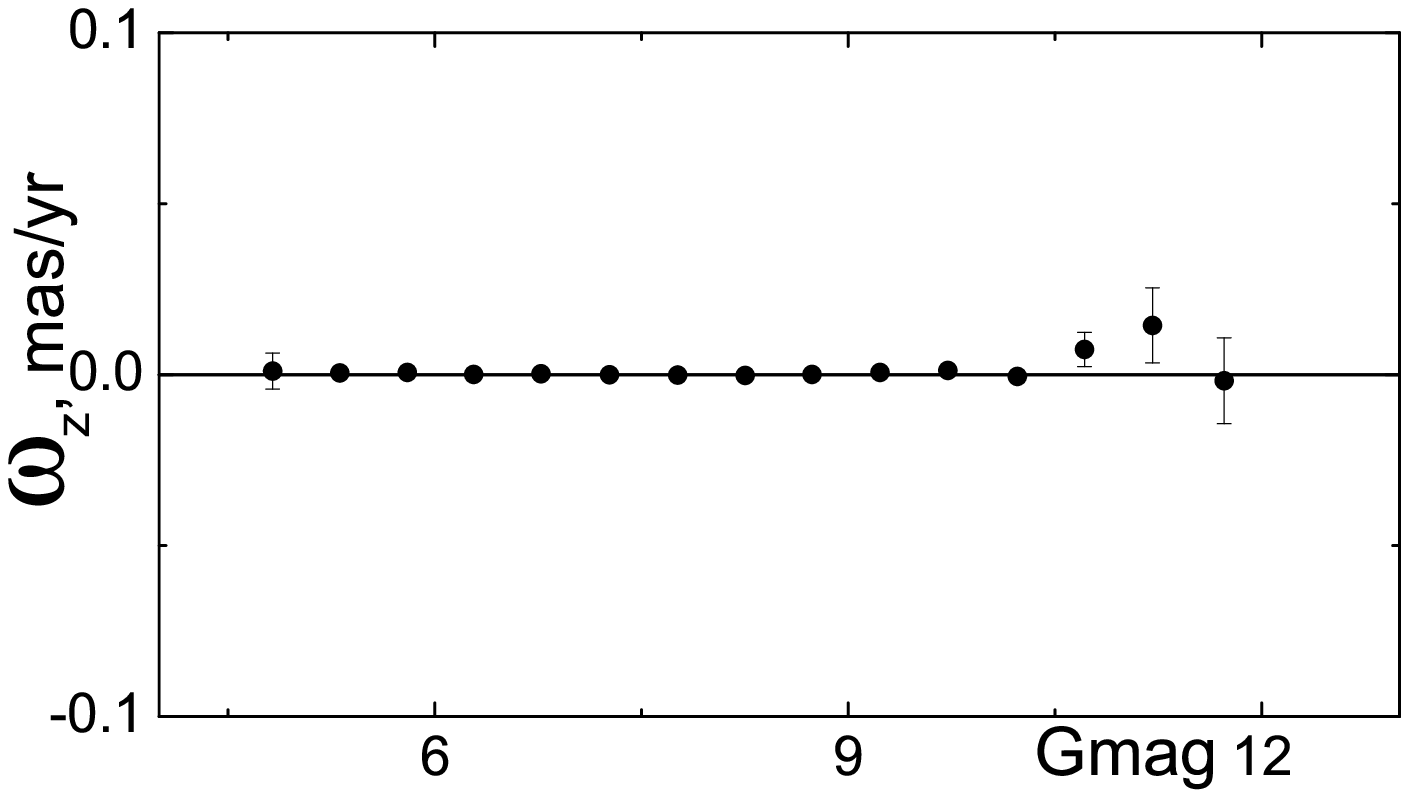}

\caption{Components of the mutual rotation vector between coordinate systems of the TGAS$_{origin}$ (with proper motions of Hipparcos-stars derived in AGIS) and TGAS$_{classic}$ (with proper motions of Hipparcos-stars calculated by the classical method) as a function of G magnitude.} 
\label{hohc}
\end{figure*}

Therefore it is obviously that proper motions of the TGAS stars can be derived at about the same accuracy level by the calssical procedure, i.e. as difference of (Hipparcos/Tycho-2-GAIA) positions taken at the corresponding epochs divided by the epoch difference. Such proper motions are called classical to distinguish them from original proper motions of the TGAS stars derived in AGIS. It turned out that for the Tycho-2 stars classical proper motions differ from original ones. At the same time, classical proper motions derived only for Hipparcos stars at the level of about 0.01 mas/yr do not differ from original ones. In figures \ref{totc}, \ref{hohc} components of mutual rotation vector derived from differences between original and classical TGAS proper motions are presented.

Also it came out that all three components of the rotational vector for the Hipparcos stars are equal to zero throughout the whole range of magnitudes. In other word, it means that the system of original proper motions set by Hipparcos stars in the $5 < m < 11$ range  coincides with that derived by traditional classical method. In contrast, as can be seen from the figure behaviour of $\omega_{y}$ for Tycho-2 stars almost exactly follows the behaviour of $\omega_{y}$ in the above mentioned figures. Hence the original TGAS proper motions of Tycho-2 stars systematically differ from classical ones. This distinction generates the $\omega_{y}$ component that varies strongly with magnitude in the range $\sim 9.5 < m < 11.5$. Out of this range it is almost constant and is $\sim$+0,5 mas/yr in the bright part of the magnitude range and $\sim$-1,5 mas/yr in the faint one. Since to derive original proper motions of the TGAS-stars the same set of AGIS procedures was applied and to derive classical proper motions of the stars the same pocedure was used, then we could not find any rational explanation for existing differences in proper motions of Tycho-2 stars and their absence in proper motions of the Hipparcos stars.

\section{ Investigating differences between classical proper motions of the TGAS and proper motions of UCAC5, HSOY and PMA.}

\begin{figure*}
\vspace*{0pt}
\includegraphics[width = 58mm,]{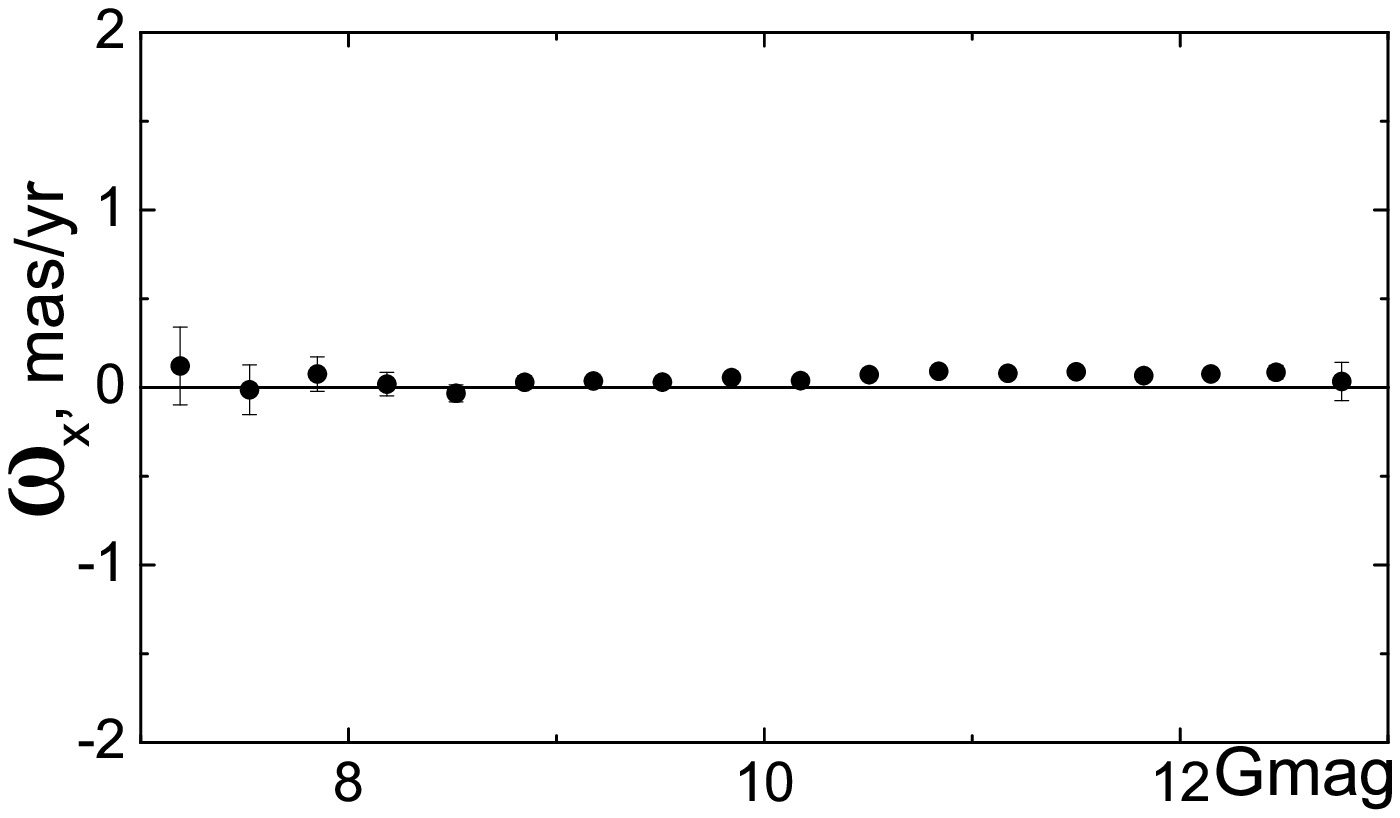}
\includegraphics[width = 58mm,]{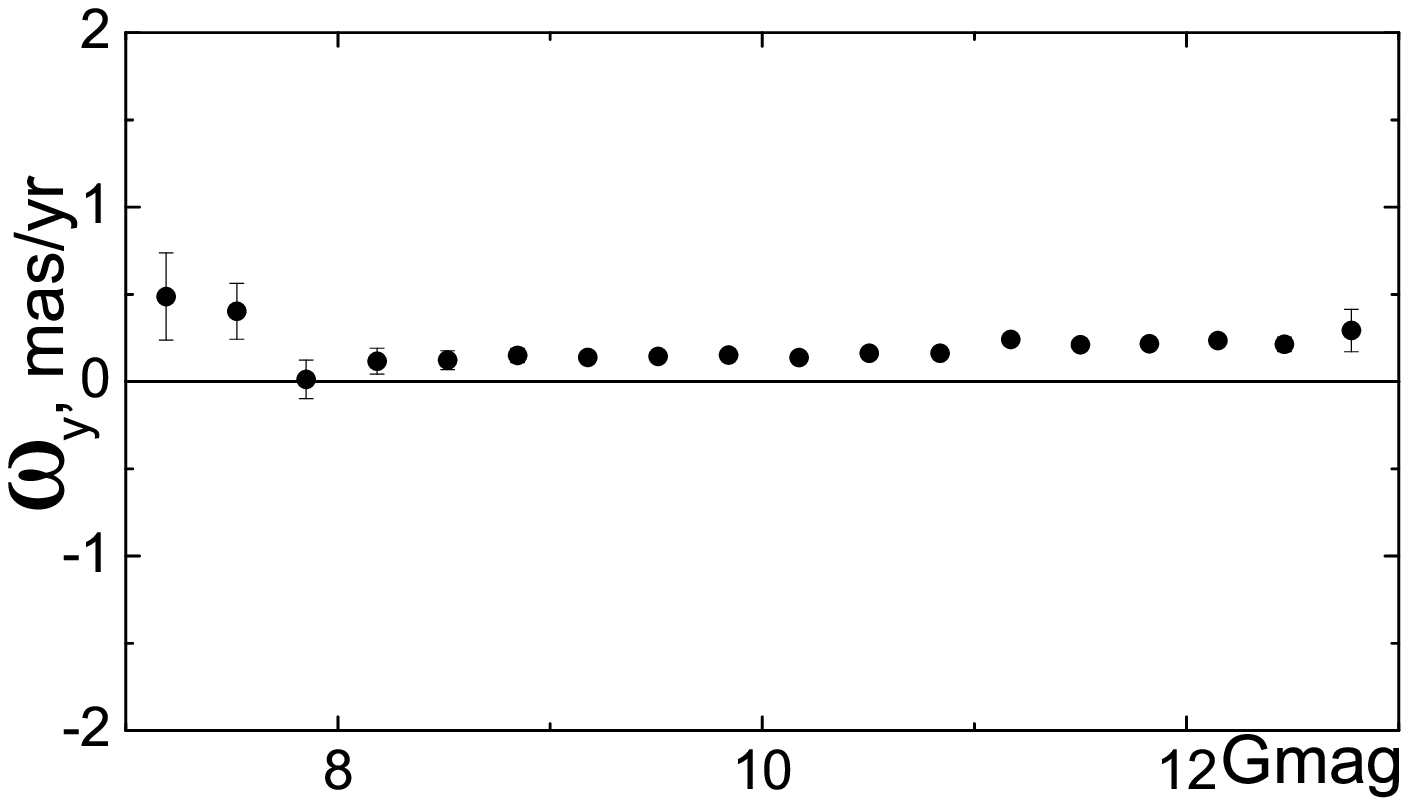}
\includegraphics[width = 58mm,]{wz_pma_tgas.eps}

\includegraphics[width = 58mm,]{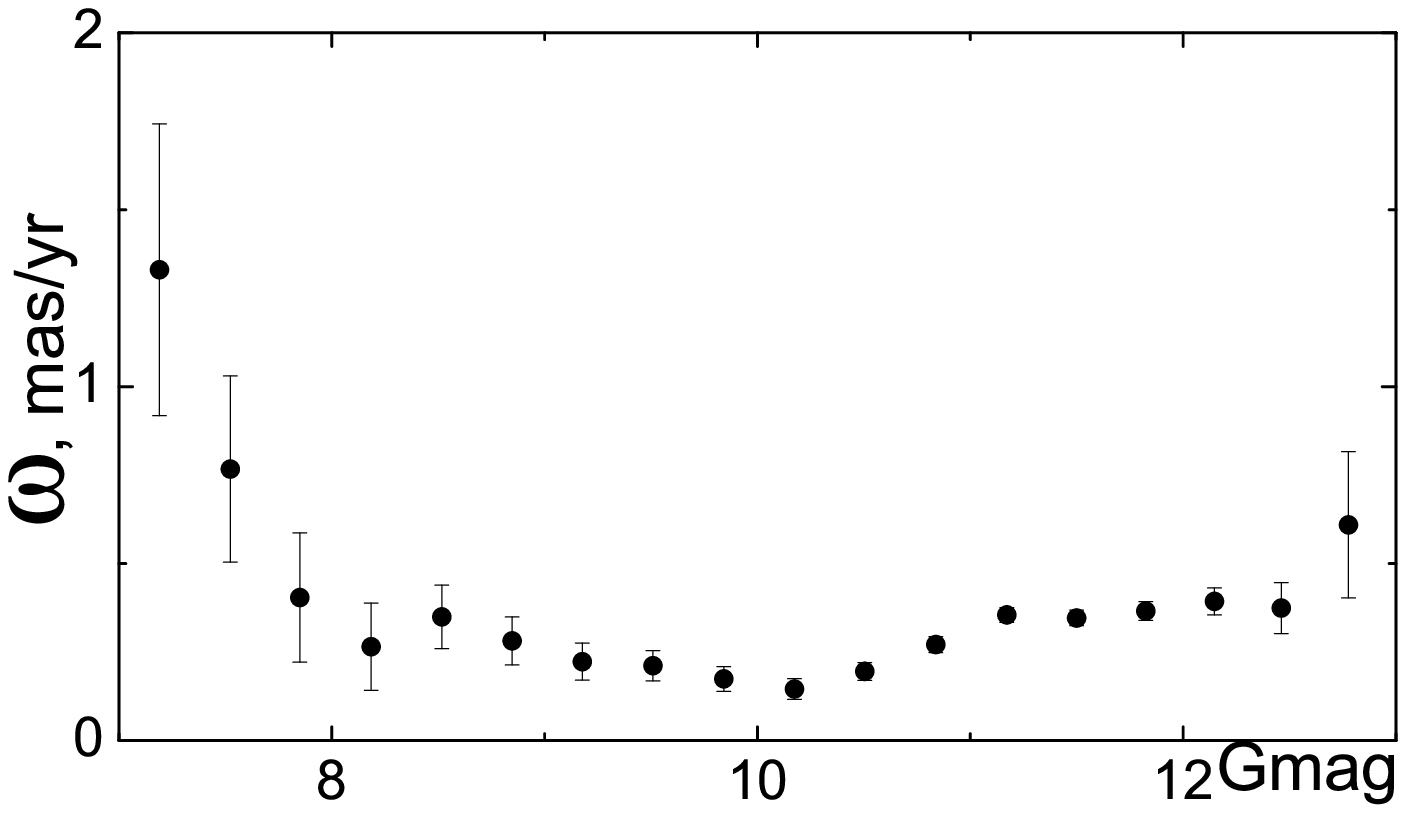}
\includegraphics[width = 58mm,]{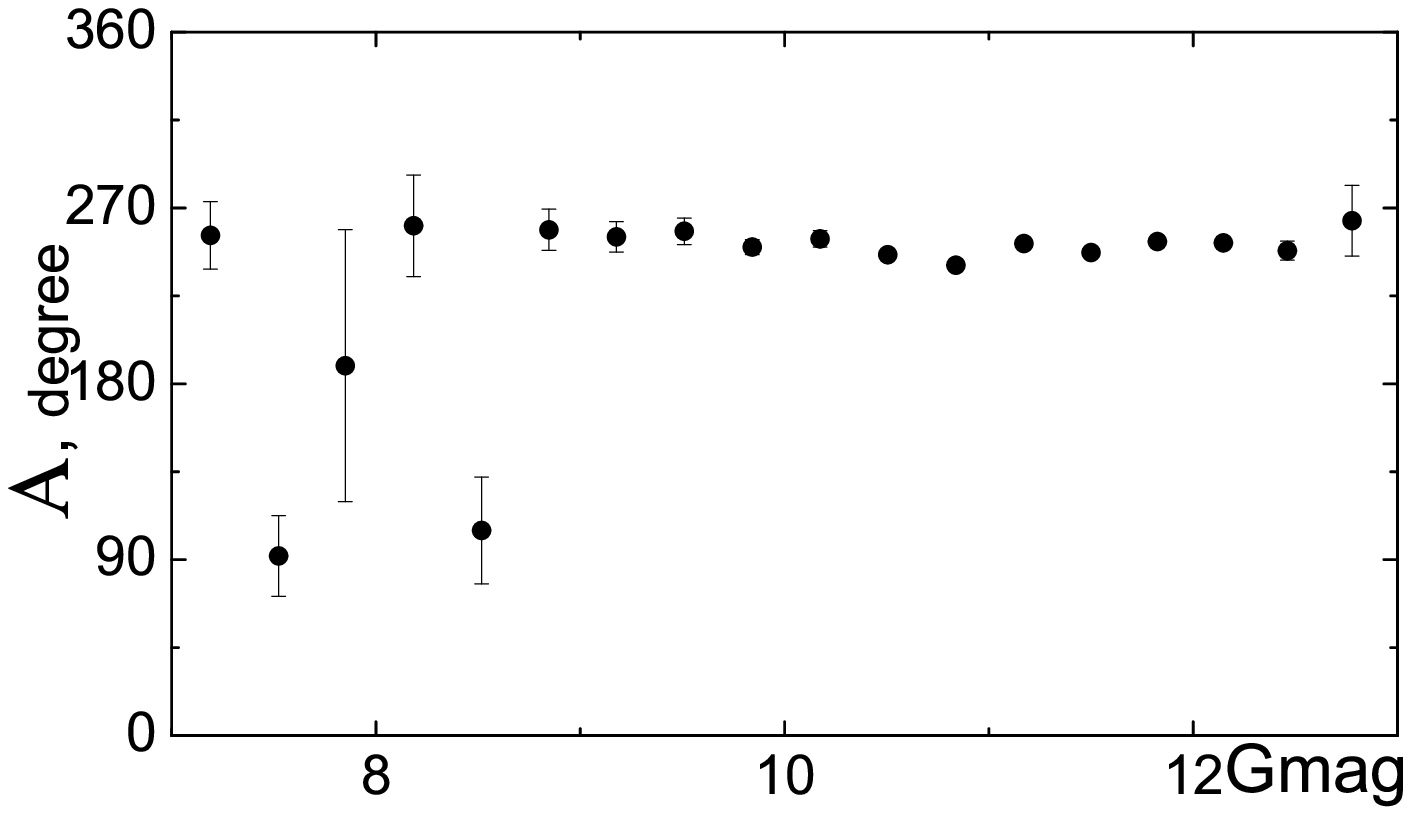}
\includegraphics[width = 58mm,]{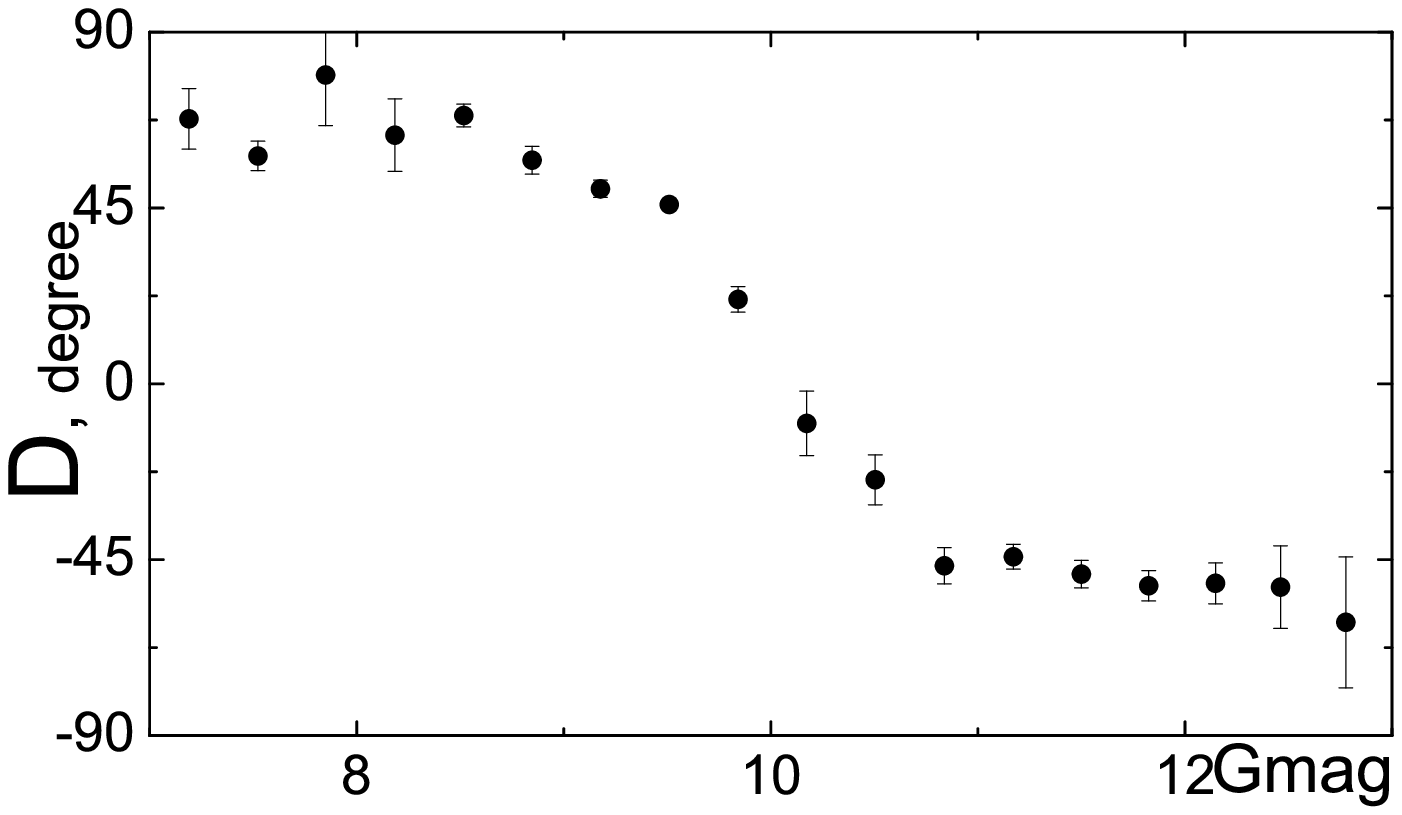}
\caption{Components of the mutual rotation vector between coordinate systems of the PMA and TGAS$_{classic}$ as a function of G magnitude (uppper panel). The modulus of mutual rotation vector and the values of pole's coordinates (bottom panel).}
\label{ptc}
\end{figure*}

\begin{figure*}
\vspace*{0pt}
\includegraphics[width = 58mm,]{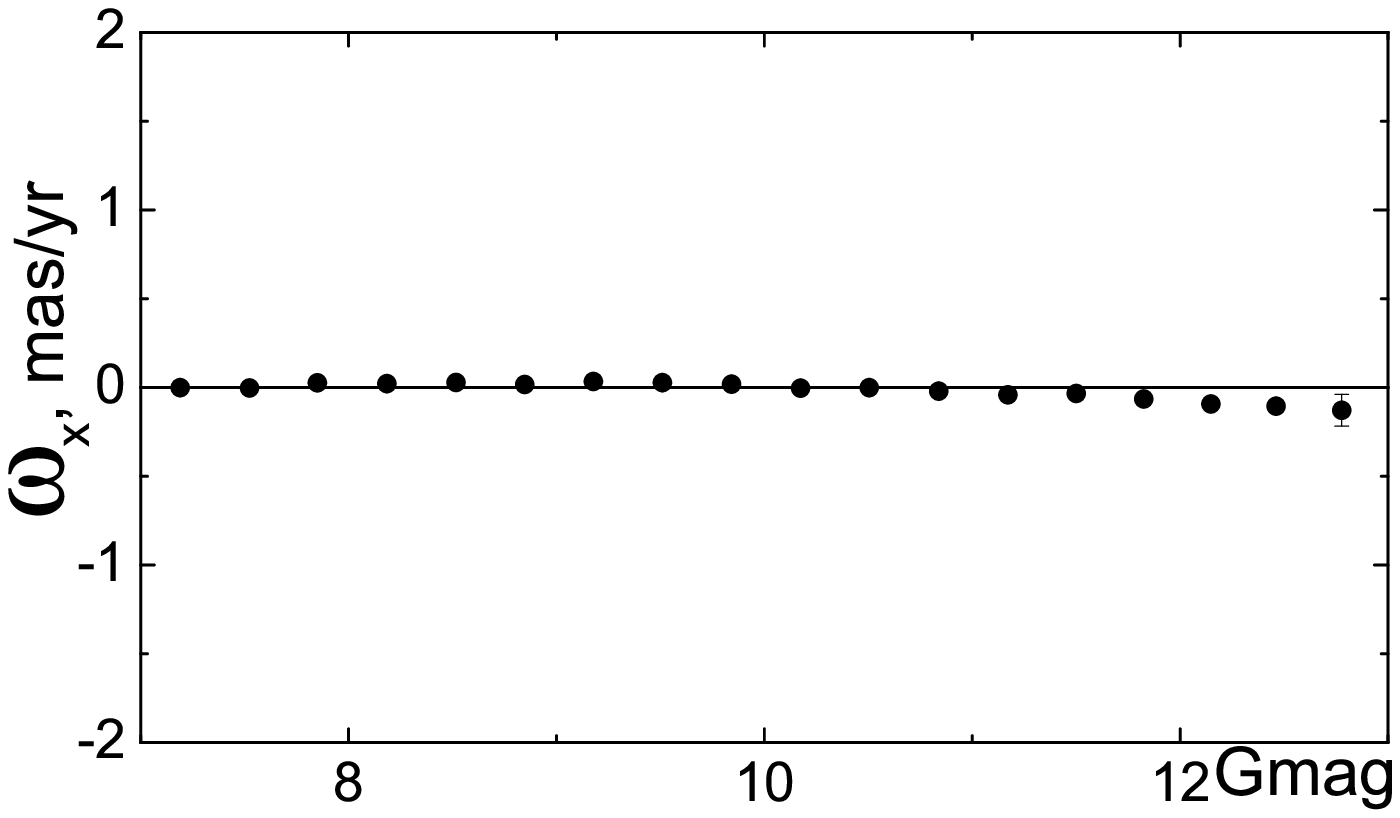}
\includegraphics[width = 58mm,]{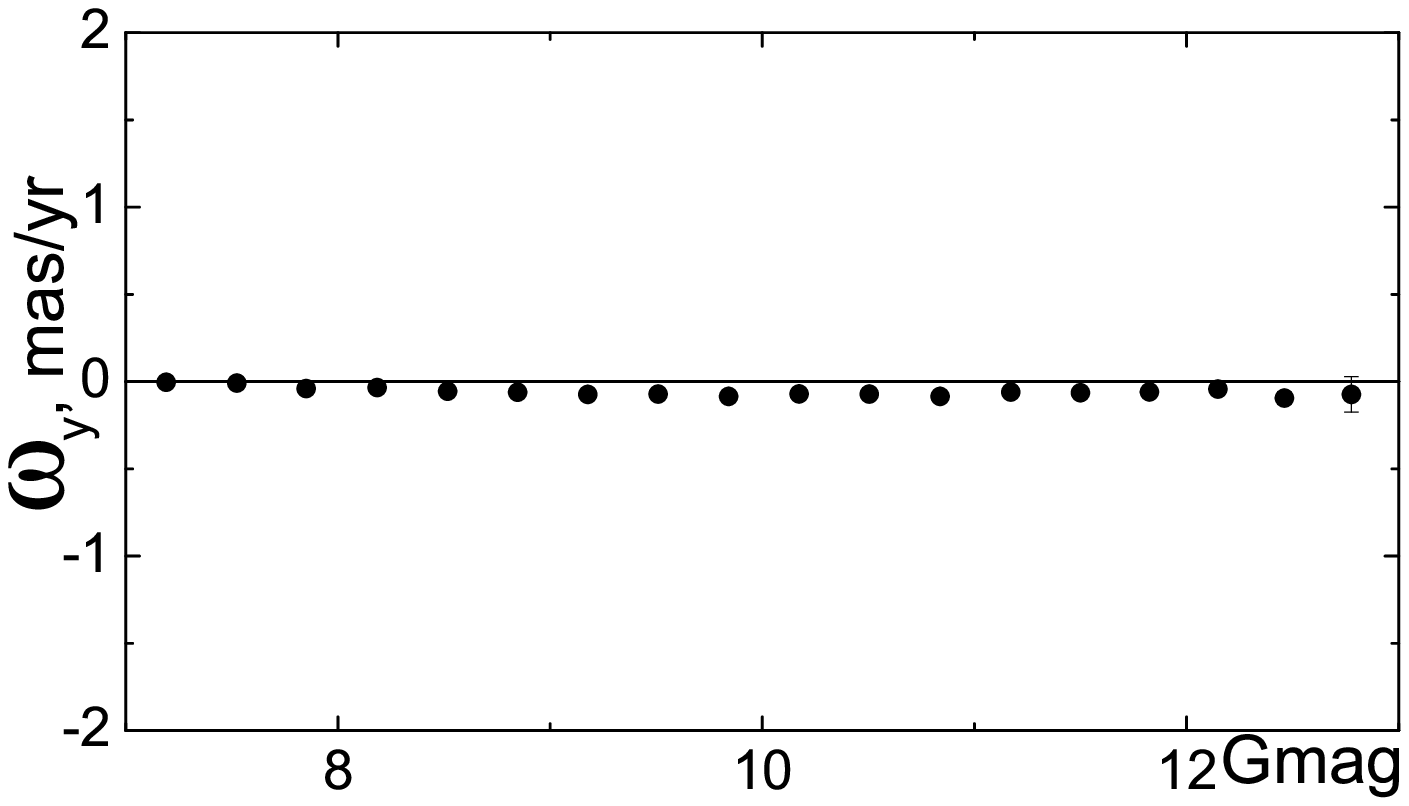}
\includegraphics[width = 58mm,]{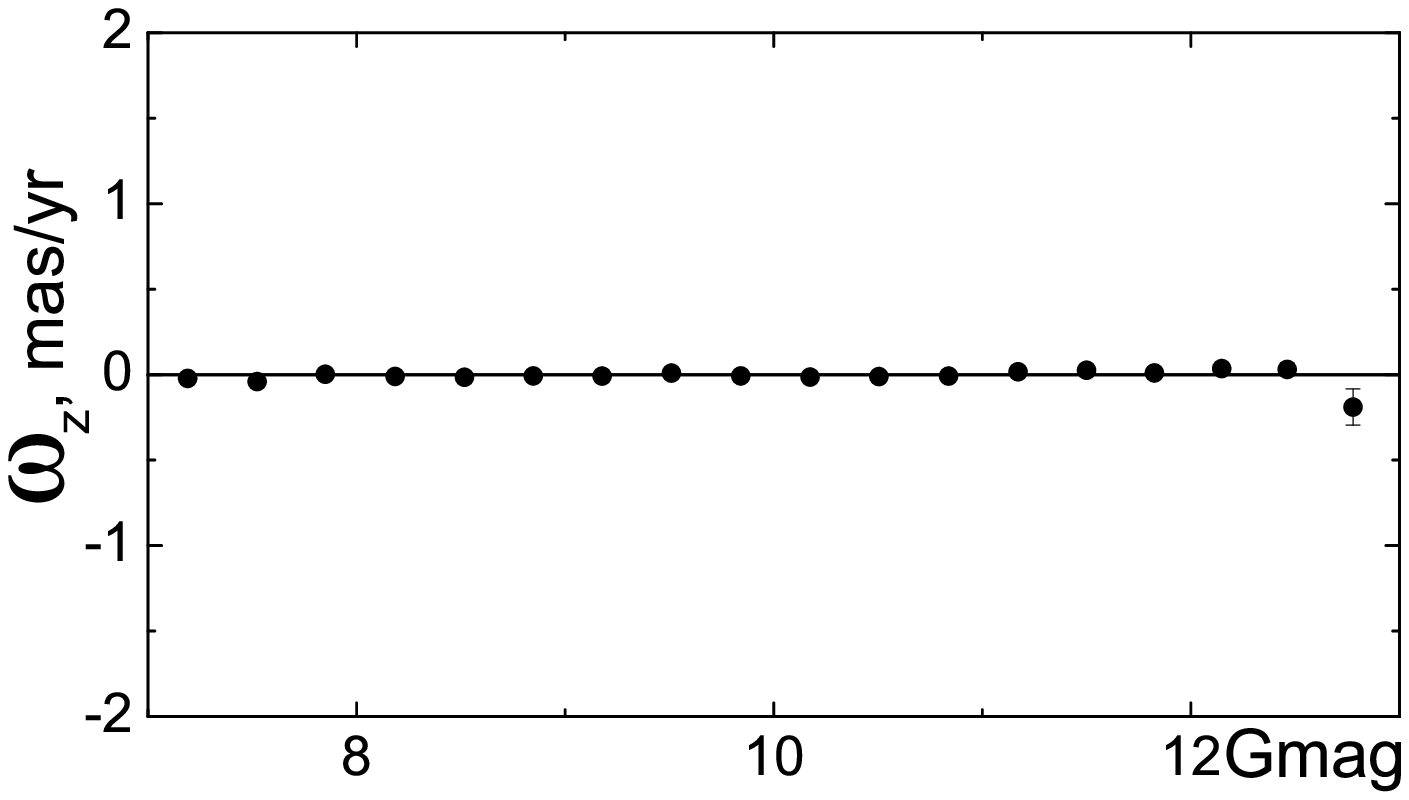}

\includegraphics[width = 58mm,]{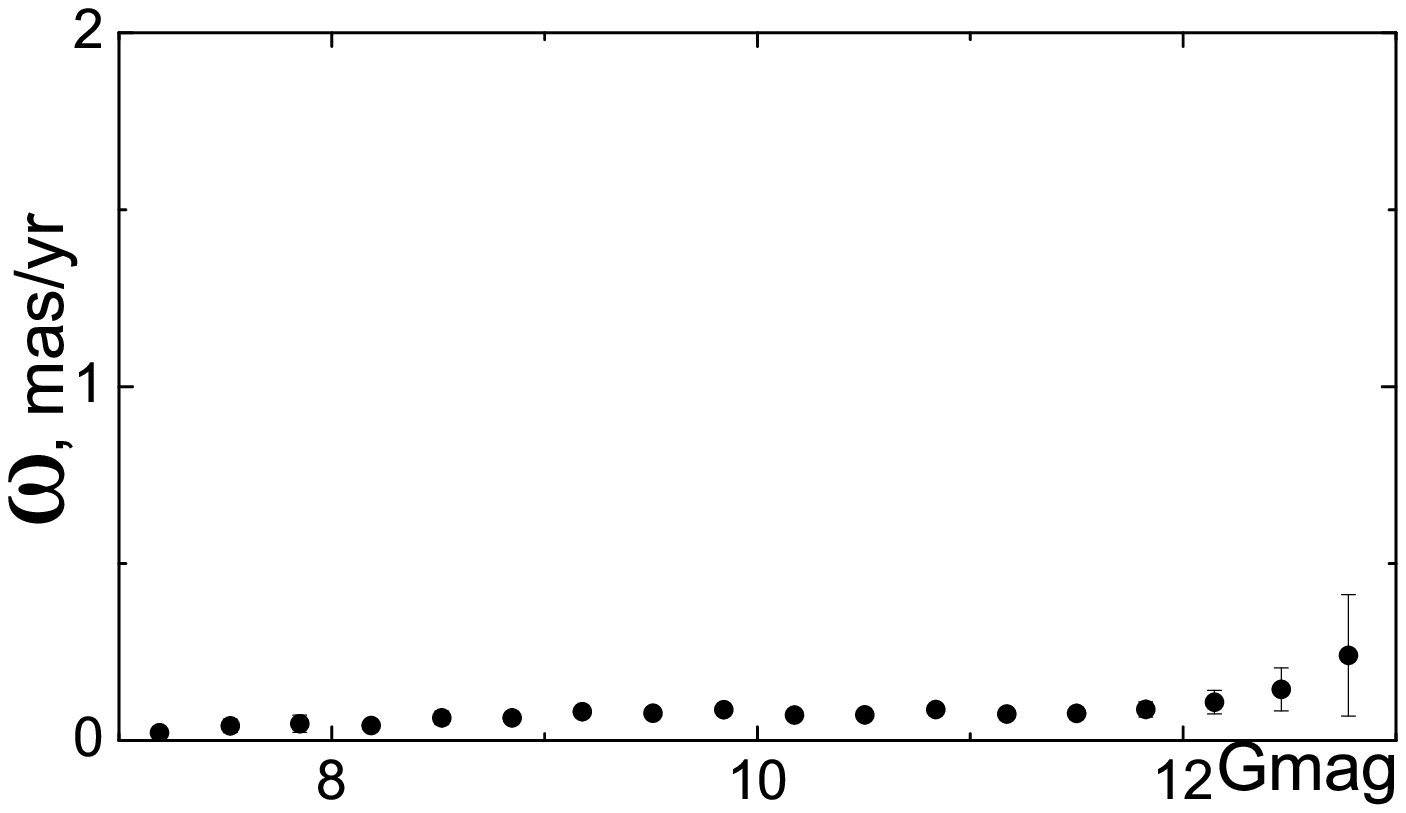}
\includegraphics[width = 58mm,]{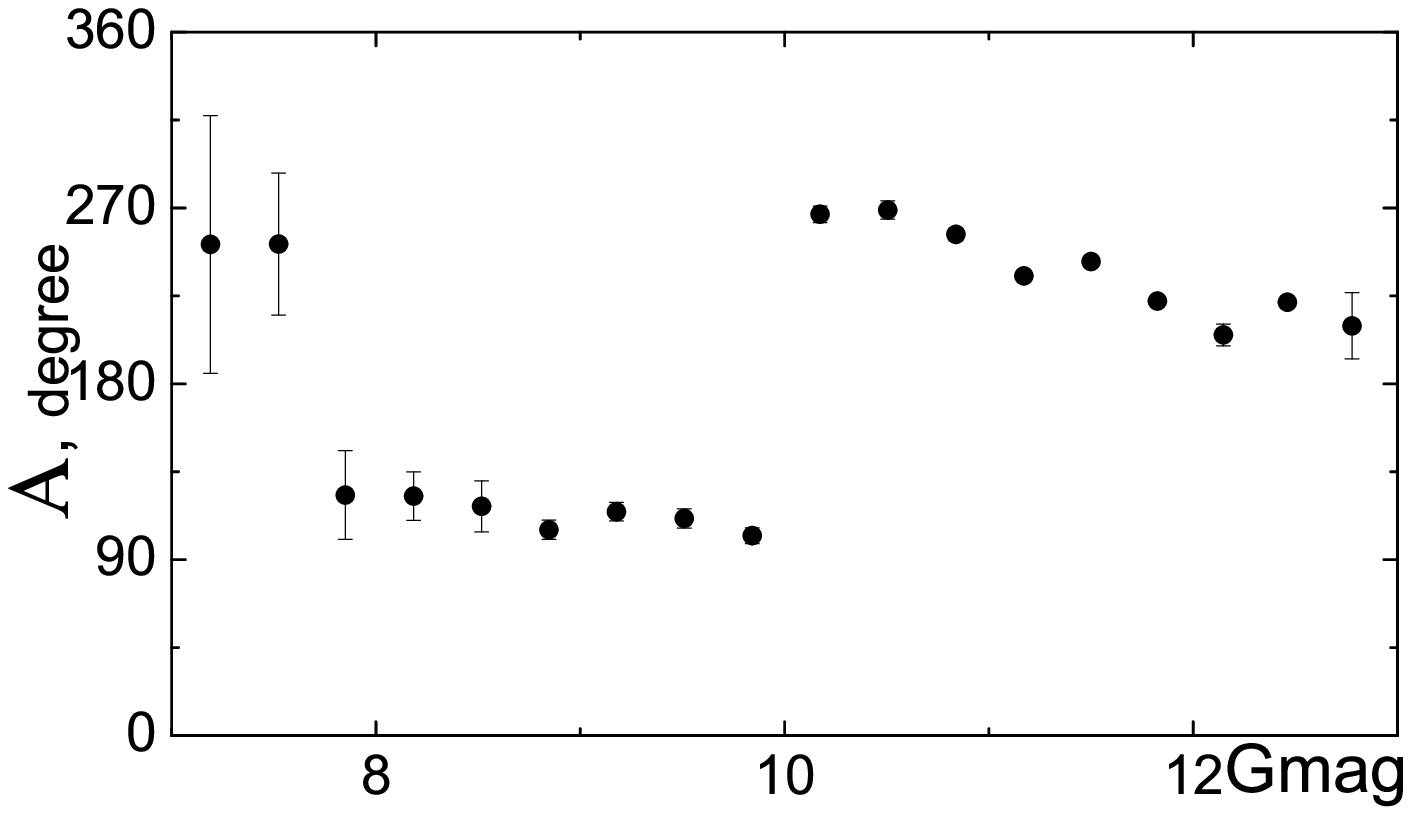}
\includegraphics[width = 58mm,]{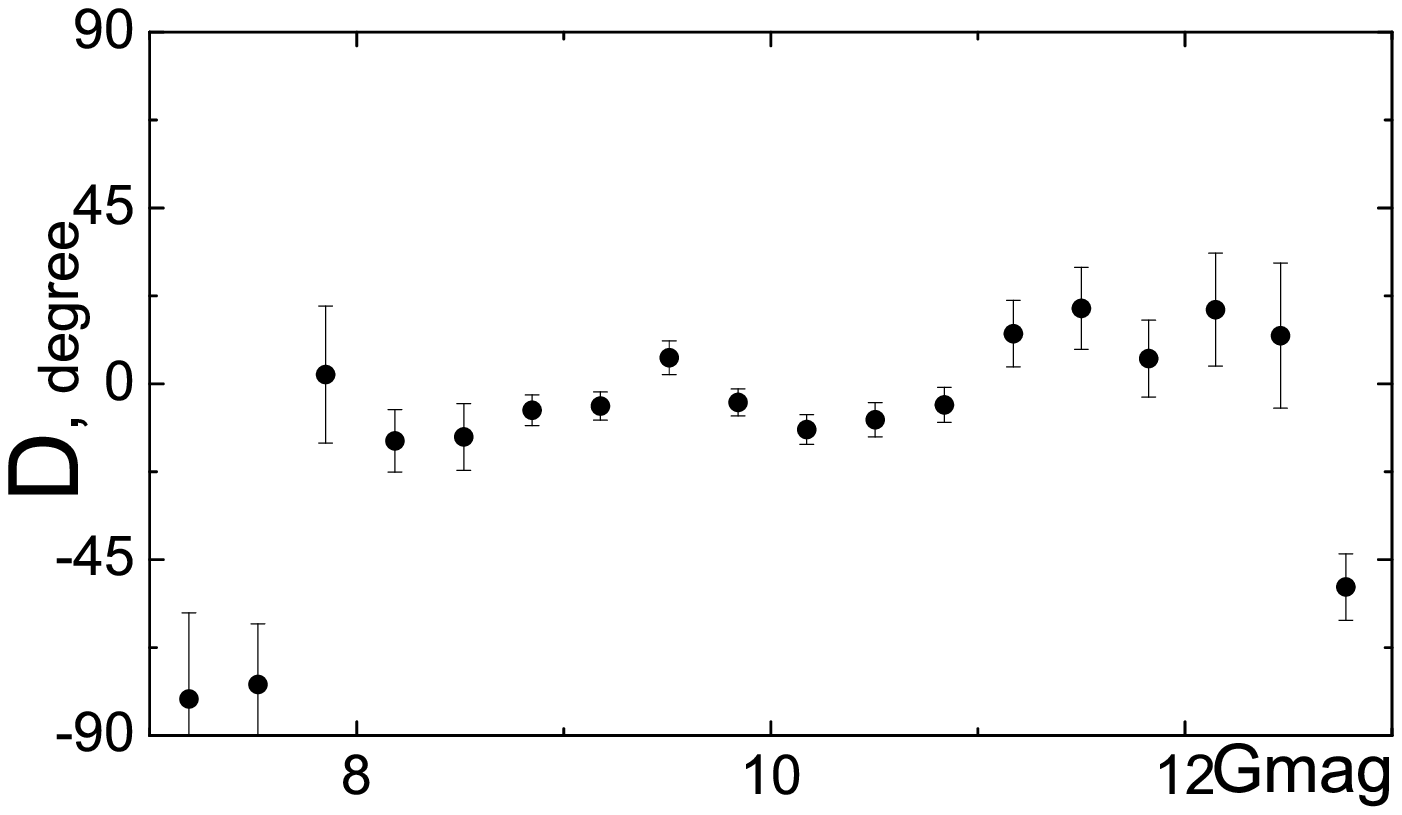}
\caption{Components of the mutual rotation vector between coordinate systems of the HSOY and TGAS$_{classic}$ as a function of G magnitude (uppper panel). The modulus of mutual rotation vector and the values of pole's coordinates (bottom panel).}
\label{htc}
\end{figure*}

\begin{figure*}
\vspace*{0pt}
\includegraphics[width = 58mm,]{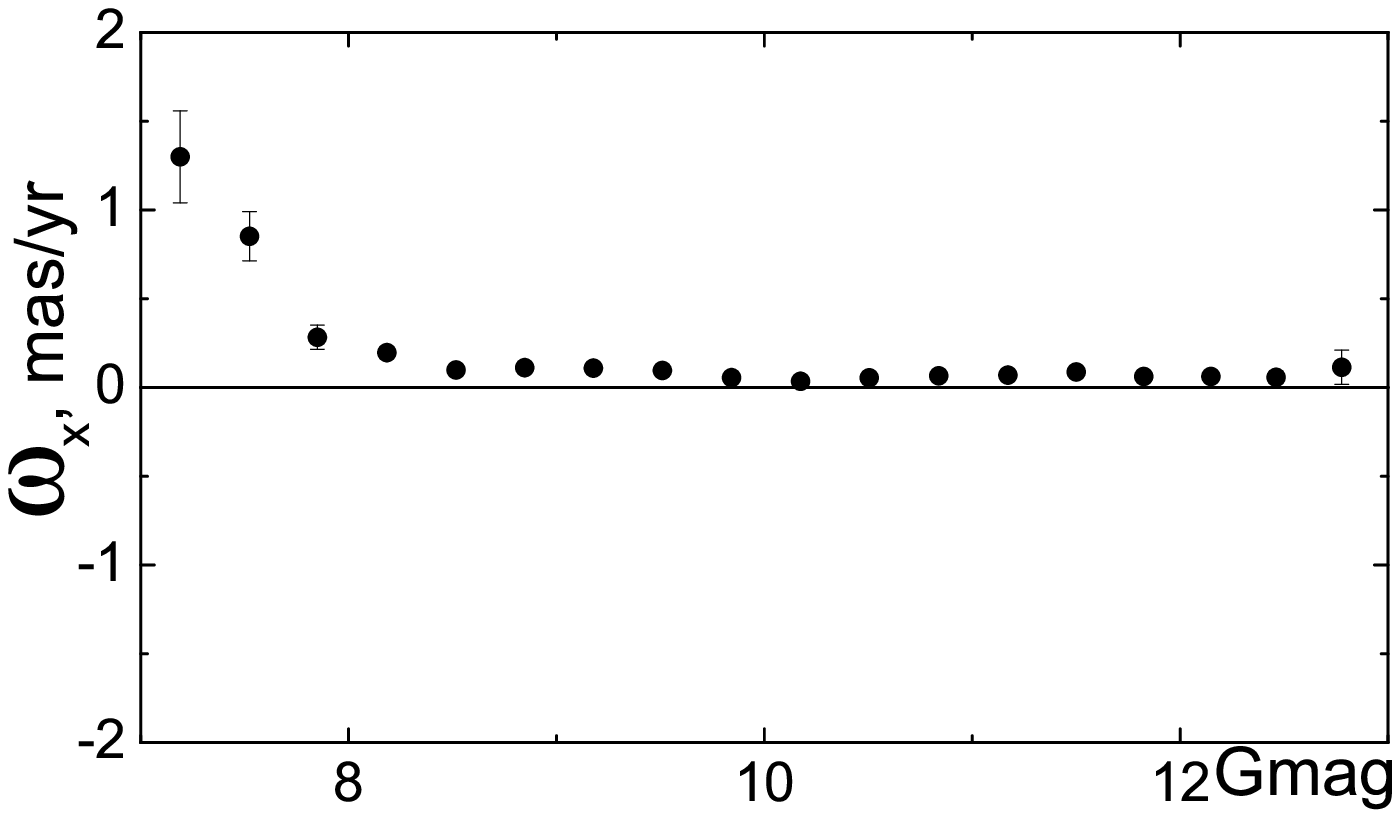}
\includegraphics[width = 58mm,]{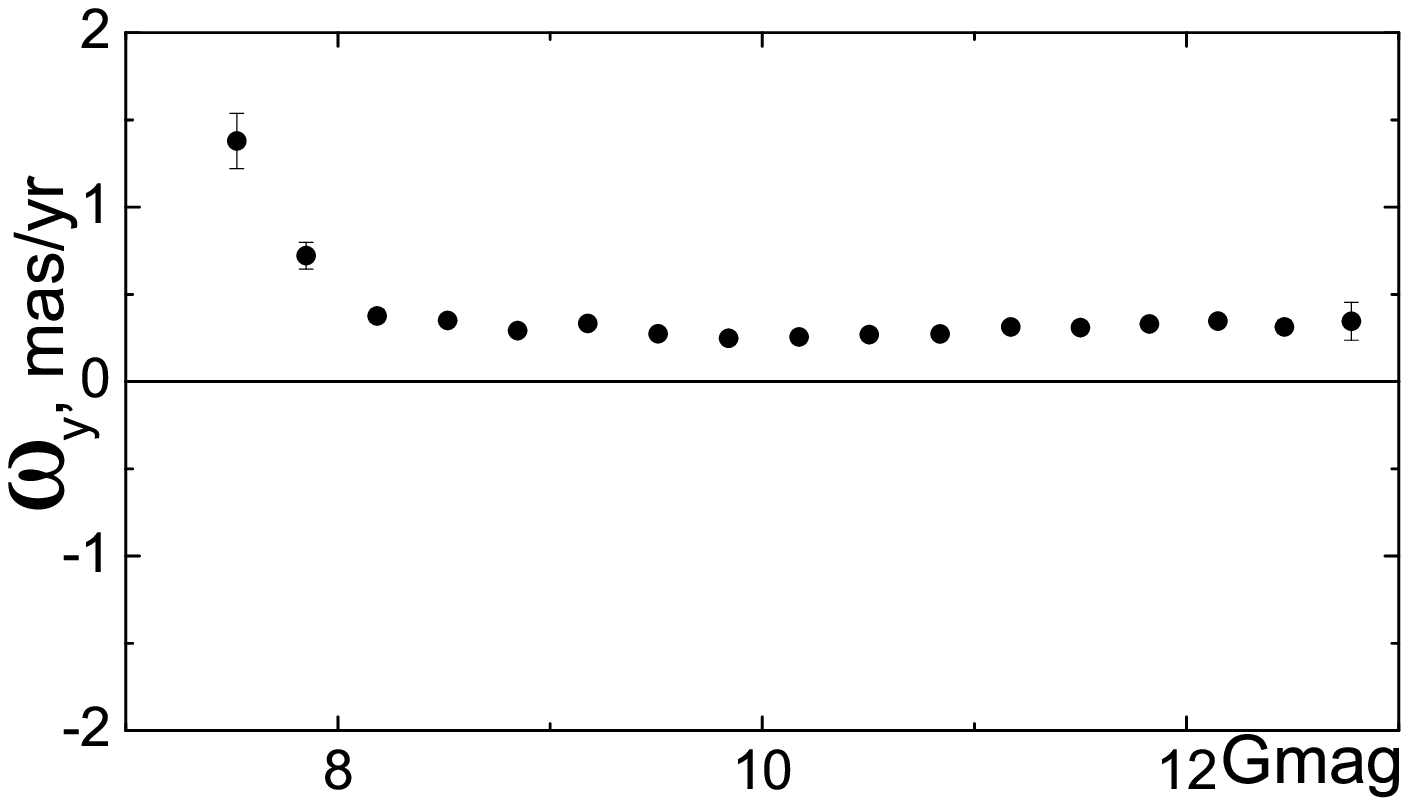}
\includegraphics[width = 58mm,]{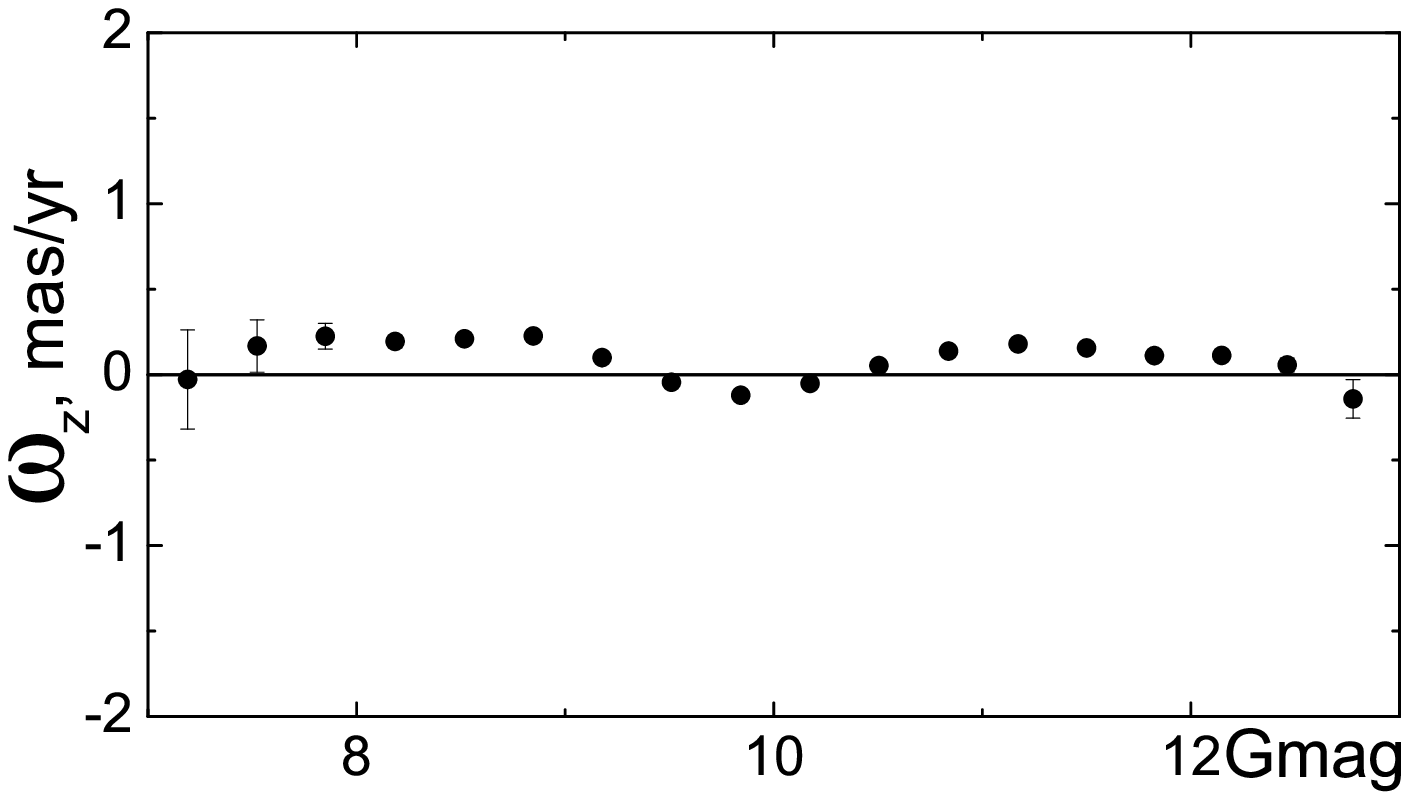}
\includegraphics[width = 58mm,]{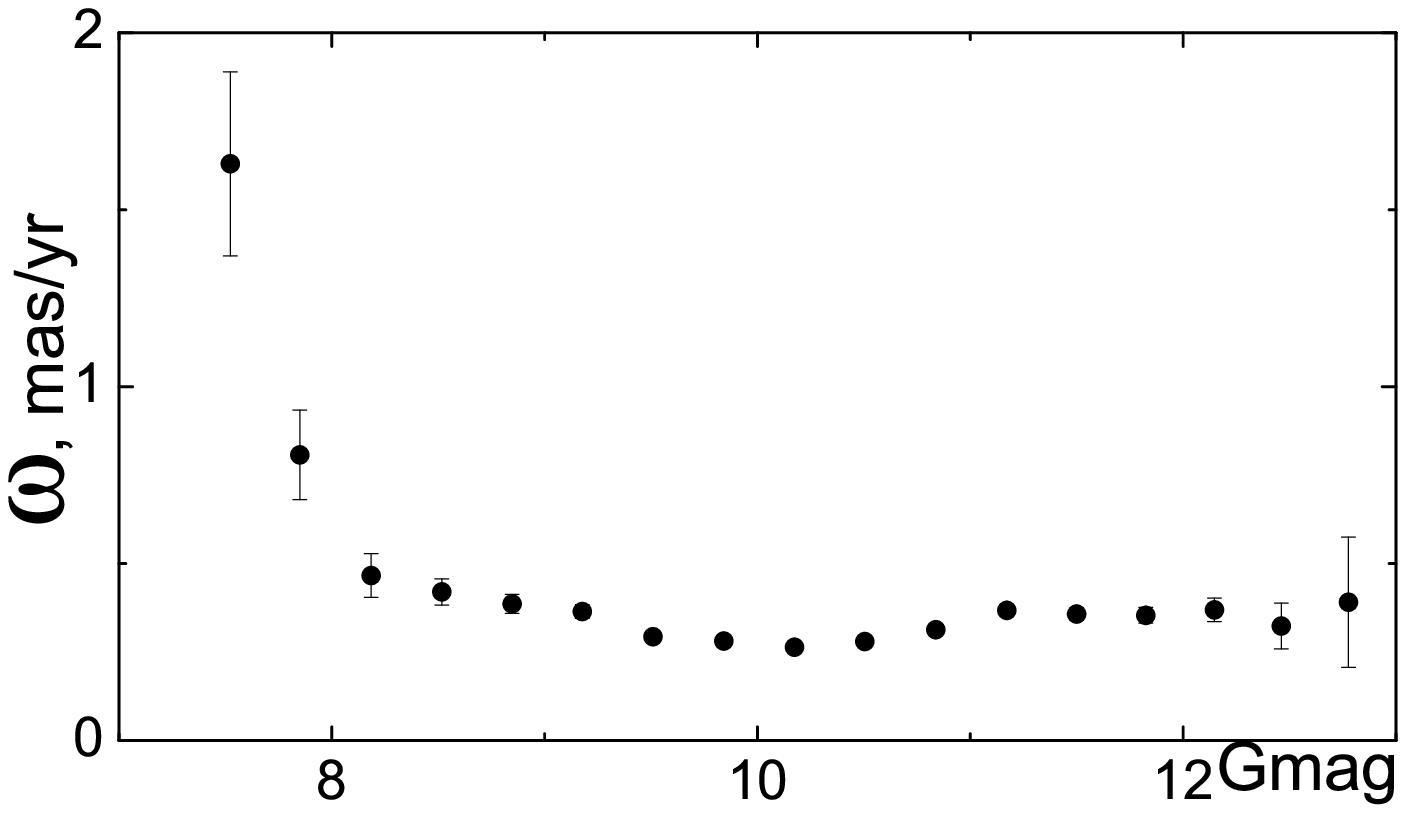}
\includegraphics[width = 58mm,]{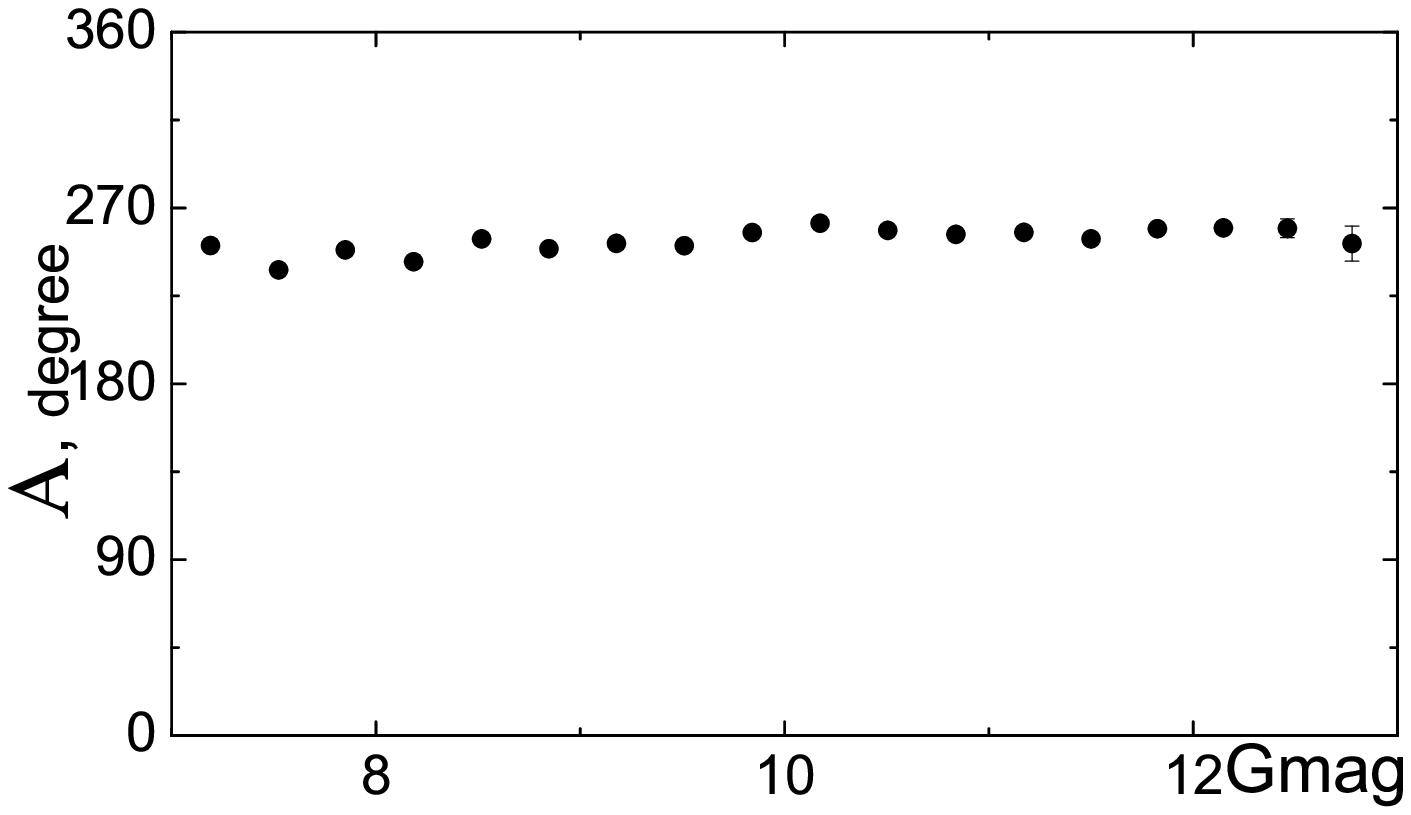}
\includegraphics[width = 58mm,]{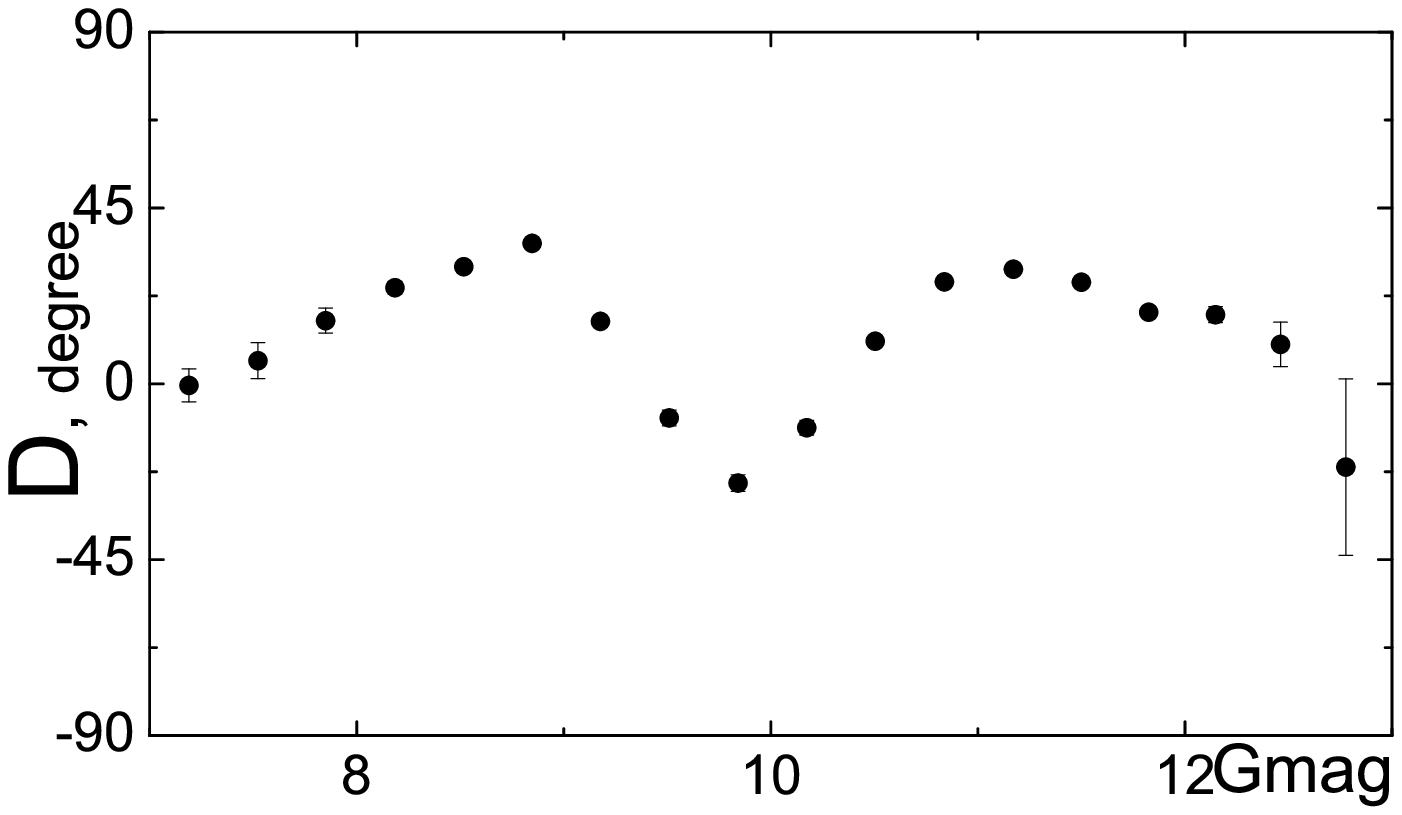}
\caption{ Components of the mutual rotation vector between coordinate systems of the UCAC5 and TGAS$_{classic}$ as a function of G magnitude (uppper panel). The modulus of mutual rotation vector and the values of pole's coordinates (bottom panel).}
\label{utc}
\end{figure*}

We compare now the proper motions of the TGAS stars calculated by the classical method with stellar proper motions of the UCAC5, HSOY and PMA catalogues. In figures \ref{ptc}, \ref{htc}, \ref{utc} dependencies of rotation vector components on magnitude are presented. We also show in the figures \ref{ptc}, \ref{htc}, \ref{utc} values of pole's coordinates where the angular velocity vector of mutual rotation is directed as well as the velocity modulus calculated using the following formula

\begin{equation}
A_{rot} = arctg \frac{\omega_{y}}{\omega_{x}}
\end{equation}

\begin{equation}
B_{rot} = arctg \frac{\omega_{z}} {\sqrt {\omega_{x}^2+\omega_{y}^2}}
\end{equation}

\begin{equation}
\Omega_{rot} =\sqrt{\omega_{x}^2+\omega_{y}^2+\omega_{z}}
\end{equation}

As can be seen from the figures, when comparing the HSOY and TGAS$_{class}$ catalogues any dependence of the rotation vector components on magnitude does not exist, and their values are virtually equal to zero. The modulus of the angular rotation velocity does not exceed 0.1 mas/yr. At the same time the $A$ coordinate of the rotation pole changes from -60$^\circ$ to -120$^\circ$ while the $D$ coordinate changes from -15$^\circ$ to +15$^\circ$ with increasing magnitude. The results of comparison of the UCAC5 and TGAS are somewhat different. $\omega_x$ and $\omega_y$ components also do not depend on magnitude albeit $\omega_y$ differs from zero by about +0.3 mas/yr. At the same time $\omega_z$ changes nonlinearly within $\pm$0.2 mas/yr although in average in the range $7 < m < 13$ magnitude it can be considered equal to zero. Modulus of angular rotational velocity in average is equal to about 0.4 mas/yr. The $D$ coordinate of the rotation pole changes almost by 30$^\circ$ in the magnitude range $7 < m < 13$ while the $A$ coordinate remains nearly constant and equal to about 80$^\circ$. The results of comparison of the PMA and TGAS catalogues are almost similar. $\omega_x$ and $\omega_y$ components do not depend on magnitude while their values do not exceed +0.25 mas/yr. The $\omega_z$ shows insignificant dependence on magnitude with a linear slope coefficient of about 0.2 mas/m. Modulus of the angular rotation velocity in average is near 0.3 mas/yr. The $A$ coordinate of the rotational pole practically does not depend on magnitude and in average $А$ $\sim$ 70$^\circ$. The $D$ coordinate of the pole depends on magnitude and changes from $D$ $\sim$ 70$^\circ$ to $D$ $\sim$ 60$^\circ$.

It is now clear that the reason for the dependency of the $\omega_y$ on magnitude discovered in section 2 is connected with the original TGAS proper motions because their replacement by classical ones leads to vanishing this dependency. At the same time the specified replacement virtually does not influence on behaviour of the $\omega_x$ and $\omega_z$. Behaviour of $\omega_z$ component when comparing UCAC5 and PMA with TGAS differs and still depends on magnitude in the narrow magnitude range from 9.5 to 11.5. This is confirmed by the dependence of the coordinates of the poles on magnitude. Values of right accention $A$ of the rotation poles between the coordinate systems (PMA-TGAS) and (UCAC5-TGAS) derived using the formula (3) are almost constant and do not depend on magnitude while values of declination $D$ (formula (4)) replicate behaviour of the $\omega_z$. It follows immediately that the dependence of the $\omega_z$  component is a consequence of dependence on magnitude containing in differences between stellar proper motions  $\Delta\mu_\alpha cos\delta$. Based on the available data it is impossible to make an unambiguous decision about in which of these catalogues under comparison the difference $\Delta\mu_\alpha cos\delta$ is burdened by magnitude error. Although it can be seen that for all the three catalogues- PMA, UCAC5 and HSOY — it is the magnitude range from  9.5 to 11.5 that is specific, and dramatic changes of the $\omega_z$ component are within it while out of this range dependence on magnitude is weak.

\begin{figure*}
\vspace*{0pt}
\includegraphics[width = 58mm,]{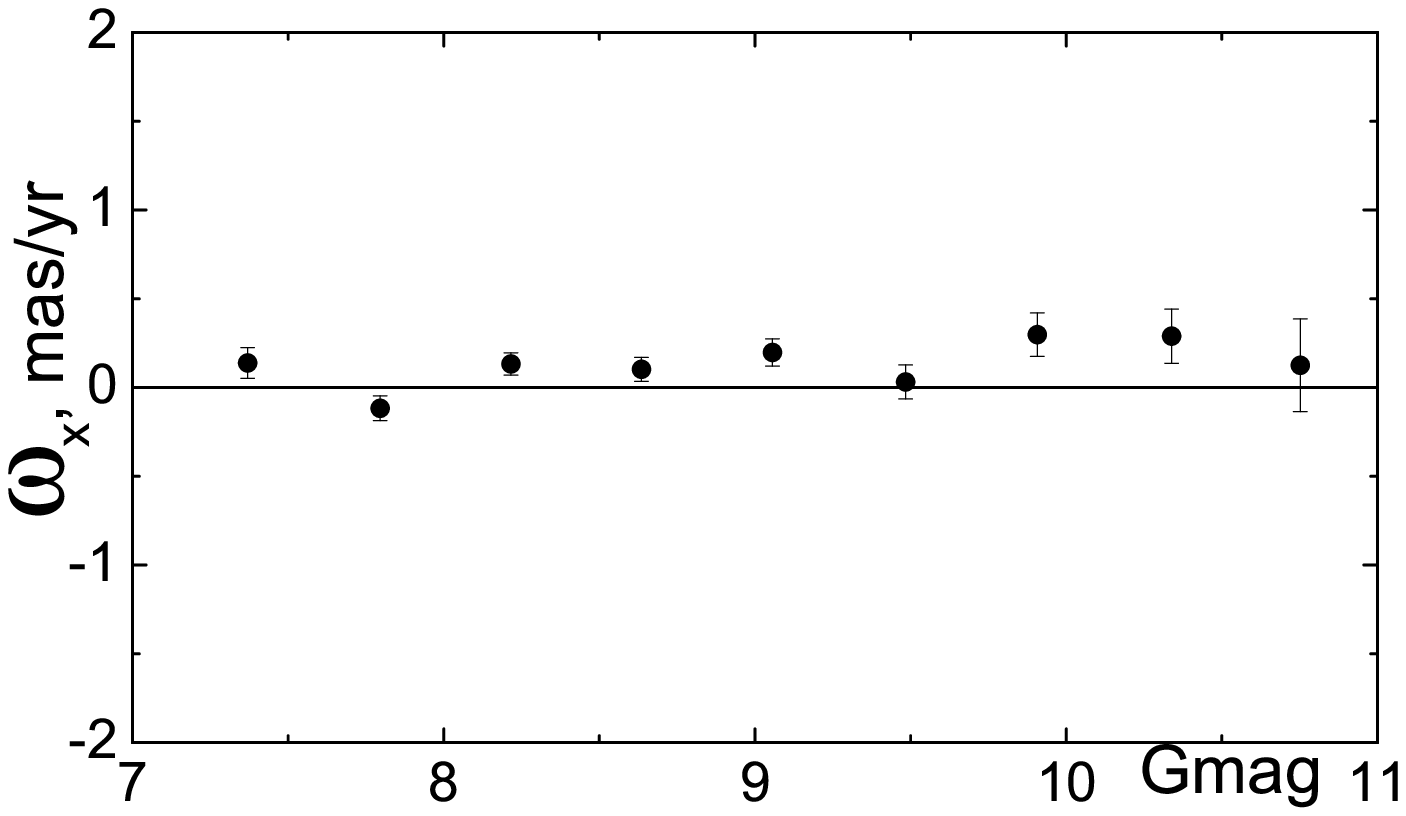}
\includegraphics[width = 58mm,]{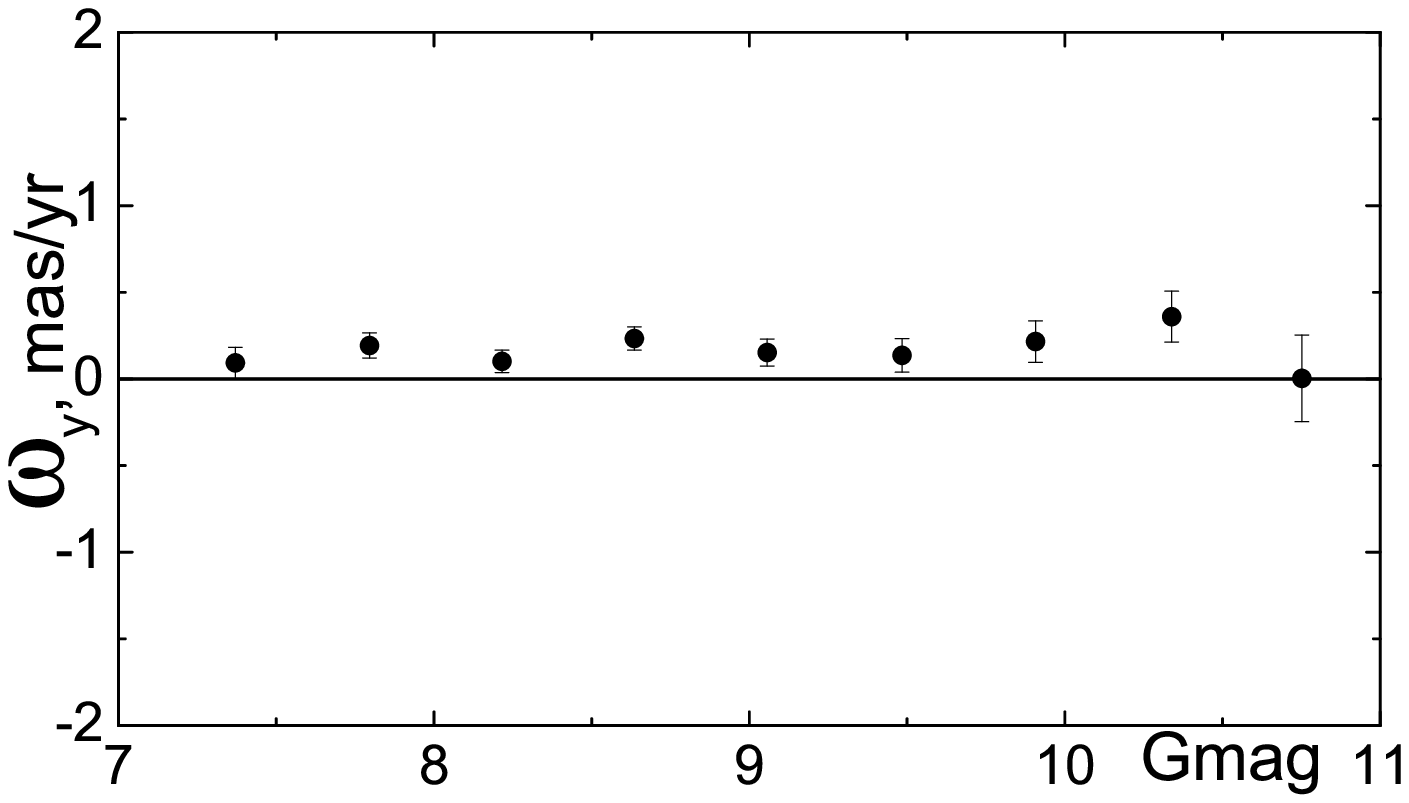}
\includegraphics[width = 58mm,]{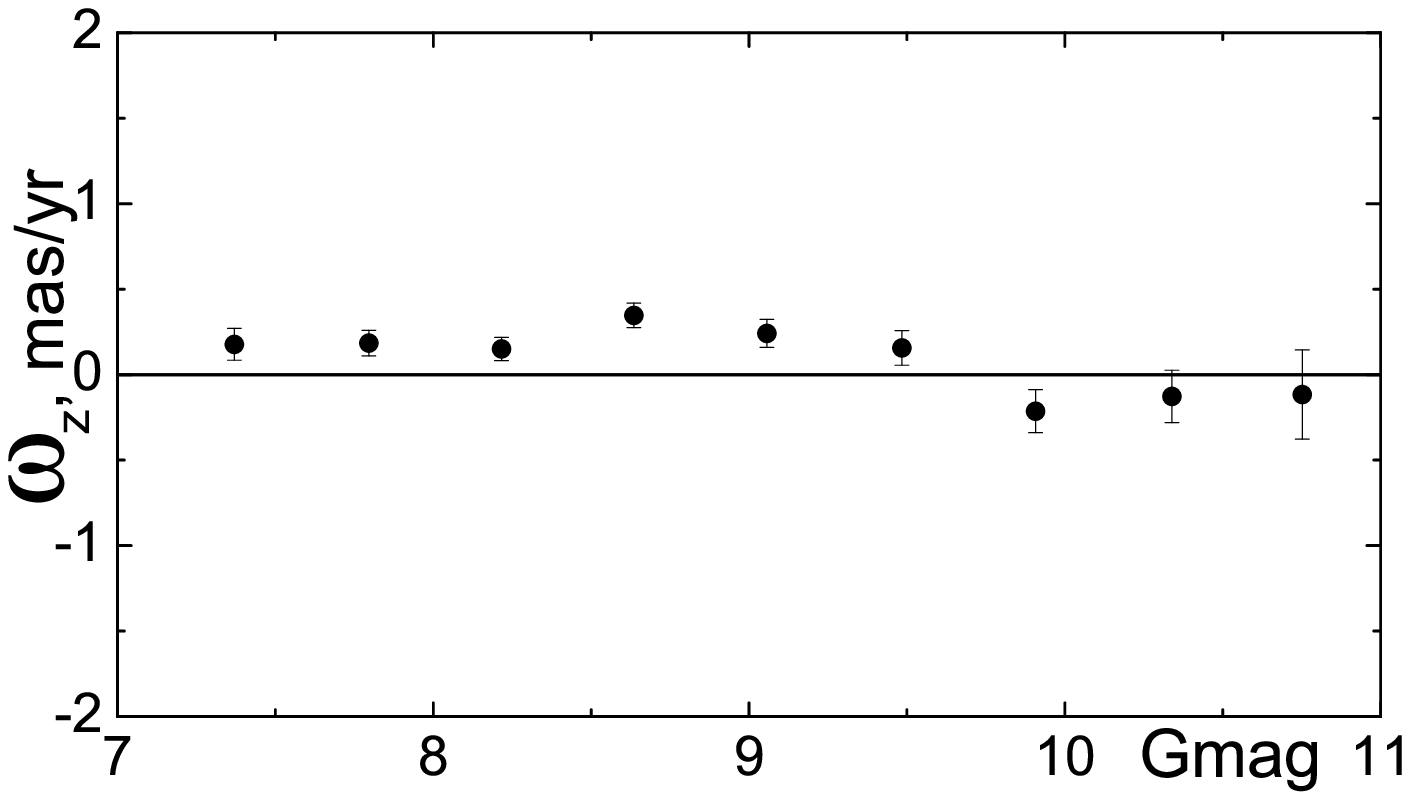}
\caption{Components of the mutual rotation vector between coordinate systems of the TGAS $_{origin}$ (only Hipparcos-stars) and PMA as a function of G magnitude.} 
\label{hphu}
\end{figure*}

Also when comparing stellar proper motions of the only Hipparcos-stars containing in the PMA with those from the TGAS, it was found out that in the range $7.75 < m <9 .75$ values of all the tree components do not exceed $\pm$ 0.25 mas/yr while dependence of the $\omega_z$ component on magnitude is determined unreliable (see Fig.\ref{hphu}). It means that in this range of magnitudes proper motions of Hipparos-stars in the PMA are close  between themselves and are almost free of any magnitude equation. This result is not surprising for the UCAC5 and HSOY because their authors sought to reproduce the reference system of the TGAS catalogue. However, it is still misterious why after reduction to the system of a reference catalogue using the TGAS data systems of proper motions of the UCAC5 and HSOY reproduce well the reference system of Hipparcos-stars and reproduce poorly the one of Tycho-2 stars resulting in the magnitude-dependent $\omega_y$  component when comparing with the TGAS.

Also we could not explain the contradiction in behaviour of the $\omega_z$ component derived with the use of only Tycho-2 stars and only Hipparcos-stars when comparing the PMA and UCAC5 with TGAS. Of course it can be assumed that Hipparcos stars in the 2MASS catalogue have positions taken right from the Hipparcos  (in published papers we did not meet any information about that), then the specified contradiction for the PMA catalogue is resolved. However it remains for the UCAC5 catalogue since it is well known that positions of the Hipparcos-stars in the UCAC5 catalogue were derived from ground-based CCD observations rather than put from the Hipparcos catalogue. Therefore, the final answer to the question about whether the dependence $\omega_z$ on  magnitude is generated by the proper motions $\mu_\alpha cos\delta$ of the PMA or TGAS catalogues, is still unknown.

\section{Conclusions}

We carried out the analysis of the three catalogues - HSOY, UCAC5 and PMA - derived in the post-Gaia period by comparing their stellar proper motions with those of the TGAS catalogue. The analysis was performed under the assumption that the systematic differences between proper motions of common stars of the catalogues are caused by mutual rigid-body rotation of the coordinate systems under consideration. It was shown that the coordinate systems realized by the TGAS catalogue and HSOY, UCAC5 and PMA catalogues respectively do not have any mutual rotations around the X and Y axes exceeding 0.5 mas/yr. At the same time, in all three cases comparison with the TGAS catalogue in the range from 9.5$^m$ to 11.5$^m$ the $\omega_y$ component demonstrates almost identical variations from +0.5 mas/yr to -1.5 mas/yr. It is clear that in this case the $\omega_y$ component cannot be interpreted as constituent of the rigid-body rotation vector. Therefore, we have assumed that the existence of the identical variations of 
the $\omega_y$ is due to systematic error containing in proper motions of the TGAS stars.
It has been shown that the original stellar proper motions of the TGAS catalogue contain systematic error which resulting in the difference between proper motions derived in AGIS procedure and those calculated by the classical method. Also we show that there is no any difference between original and classical proper motions of Hipparcos stars in the TGAS catalogue up to hundreds of milliarcsecond. It unambiguously indicates that the original proper motions of Tycho-2 stars are burdened by a systematic error. Explanation of the origin of the discovered error turned out to be a back-breaking task.
Using the classical proper motions for TGAS stars, we first of all note that the angular velocity modules of mutual rotation in the range from 9.5$^m$ to 11.5$^m$ have a pronounced singularity, although their average value in the whole magnitude range of TGAS, can be considered as a measure of the residual rotation, and does not exceed 0.4 mas/yr. 
In framework of the adopted solid-body model the coordinate systems set by the TGAS catalogue with classical proper motions as well as by the HSOY, UCAC5 and PMA catalogues have mutual rotation the components of which are less than when using original proper motions of the TGAS and do not exceed 0.2-0.3 mas/yr.

Finally, we demonstrate that stellar proper motions of the PMA do not have magnitude equation when comparing them with Hipparcos stars containing in the TGAS. The contradiction related to the absence of the magnitude equation in the PMA when considering the Hipparcos stars and its existence when considering Tycho-2 stars remains unresolved.
The errors found in the proper motions of the TGAS likely have to correlate with values of the parallaxes the need for corrections of which is indicated in papers (see, for example, \citep{Astraa2016a, DeRidder2016,Davies2017}). It is clear that the values of the errors discovered are not negligible and undoubtedly have to be eliminated. Nevertheless, we consider the TGAS catalogue first of all as the useful demonstration of opportunities of the Gaia project and our paper, testing data of the TGAS catalogue, - as intention to realize fully its potential. The use of Hipparcos and Tycho-2 data in the primary solution was the forced step for a number of reasons, primarily due to short duration of the observational period. The final relese of the Gaia as was noted in \citep{l1} will be self-sufficient in the sense that any external data will not be used to derive astrometric parameters and fix the reference frame. Probably, it will provide an absence of the errors found by us if they are caused by the use of external data rather than generated in using different processing 
procedures. 

\section{ACKNOWLEDGEMENTS}

This work has made use of the data from the European Space Agency (ESA) mission Gaia (http://www.cosmos.esa.int/gaia), processed by the Gaia Data Processing and Analysis Consortium (DPAC, http://www.cosmos.esa.int/web/gaia/dpac/consortium). Funding for the DPAC has been provided by national institutions, in particular the institutions participating in the Gaia Multilateral Agreement.

\bsp

\label{lastpage}


\begin{thebibliography}{99}


\bibitem[\protect\citeauthoryear{Altmann et al.}{2017}]{a4} Altmann M., Roeser S., Demleitner M., Bastian U., Schilbach E., 2017, A\&A, 600, L4

\bibitem[\protect\citeauthoryear{Akhmetov et al.} {2017}] {a5} Akhmetov V.S., Fedorov P.N., Velichko A.B., Schulga V.M., 2017, MNRAS, 469, 763 

\bibitem[\protect\citeauthoryear{Astraatmadja \& Bailer-Jones}{2016}]{Astraa2016a} Astraatmadja T.~L., Bailer-Jones Coryn~A.~L., 2016, ApJ, 833, 119

\bibitem[\protect\citeauthoryear{Davies et al.}{2017}]{Davies2017} Davies G.~R. et al., 2017, A\& A, 598, L4

\bibitem[\protect\citeauthoryear{Fabricius et al.}{2016}]{f1} Fabricius C., Bastian U, Portell  J., Castañeda J., Davidson M.,  Hambly N.C.,  Clotet M., Biermann M., Mora A., 2016, A\&A, 393, 133

\bibitem[\protect\citeauthoryear{Fey et al.}{2015}]{f2} Fey A. L., Gordon, D., Jacobs, C. S., et al. 2015, AJ, 150, 58

\bibitem[\protect\citeauthoryear{Gaia collaboration et al.}{2016b}]{b1} Gaia collaboration, Brown A. G. A., Vallenari A., et al., 2016b, A\&A, 595, id.A2, 23 pp.

\bibitem[\protect\citeauthoryear{Gaia collaboration et al.} {2016a}]{p3} Gaia collaboration, Prusti T., de Bruidjne J.H.J., et al., 2016a, A\&A 595, id.A1, 36 pp. 

\bibitem[\protect\citeauthoryear{H\o{}g et al.}{2000}]{h1} H\o{}g E. et al., 2000, A\&A, 355, L27

\bibitem[\protect\citeauthoryear{Lindegren et al.}{2016}]{l1} Lindegren L. et al., 2016, A\&A 595, id.A4, 32 pp

\bibitem[\protect\citeauthoryear{Lindegren et al.}{2012}]{l3} Lindegren L., Lammers U., Hobbs D., O’Mullane W., Bastian U., Hernández J., 2012, A\&A, 538, A78. 
 
\bibitem[\protect\citeauthoryear{Lindegren \& Kovalevsky}{1995}]{l4} Lindegren L., Kovalevsky J., 1995, A\&A, 304, L189

\bibitem[\protect\citeauthoryear{Ma et al.}{2009}]{m5} Ma C. et al., 2009, IERS Technical Note, 35

\bibitem[\protect\citeauthoryear{Michalik, Lindegren \& Hobbs}{Michalik et al.}{2015}]{m4} Michalik D., Lindegren L., Hobbs D., 2015, A\&A, 574, A115

\bibitem[\protect\citeauthoryear{Perryman et al.} {1997}] {p1} Perryman M.A.C. et al., 1997, A\&A, 323, L49

\bibitem[\protect\citeauthoryear{De Ridder et al.}{2017}]{DeRidder2016} De Ridder J., Molenberghs G., Eyer L., Aerts C., 2016, Astron \& Astrophys, 595, L3

\bibitem[\protect\citeauthoryear{Roeser et al.}{2010}]{r1} Roeser S., Demleitner M.,  and Schilbach E., 2010, AJ, 139, 2440

\bibitem[\protect\citeauthoryear{Skrutskie et al.}{2006}]{s2} Skrutskie M.F. et al., 2006, ApJ, 131, 1163

\bibitem[\protect\citeauthoryear{Stassun \& Torres}{2017}]{Stassun2017} Stassun~K.~G., Torres~G., ApJ, submitted [arXiv:1609.05390]

\bibitem[\protect\citeauthoryear{van Leeuwen}{2007}]{l2} van Leeuwen F., 2007,  A\&A, 494, L799,

\bibitem[\protect\citeauthoryear{Zacharias et al.}{2017}]{z1} Zacharias N., Finch C.,  Frouard J., 2017, AJ, 153, 2184
	
\end{thebibliography}
\end{document}